\documentclass[12pt]{iopart} 
\usepackage{iopams}
\usepackage{graphicx}
\usepackage[caption=false]{subfig}
\usepackage{psfrag}
\usepackage{cite}

\newcommand{\allowedpairwalks}{\Omega}
\newcommand{\allowedpairwalksfull}{\Omega \left(0^{+},0^{+} \right)}

\newcommand{\alw}[2]{\allowedpairwalks \left(#1, #2 \right)}

\newcommand{\trivialwalk}{\bullet}

\newcommand{\mbcontactsites}{m_d (\varphi)}
\newcommand{\tmcontactsites}{m_c (\varphi)}

\begin{document} 
\title[Three interacting friendly walks] {An exact solution of three interacting friendly walks in the bulk\footnote{Dedicated to Professor Rodney Baxter on the occasion of his 75th birthday}}

\author{R Tabbara$^1$ and A L Owczarek$^1$ and A Rechnitzer$^2$}

\address{$^1$School of Mathematics and Statistics, The University of Melbourne, Victoria 3010, Australia.}
\address{$^2$Department of Mathematics, University of British Columbia, Vancouver V6T 1Z2, British Columbia, Canada.} 
\ead{\mailto{tabbarar@ms.unimelb.edu.au}, \mailto{owczarek@unimelb.edu.au}, \mailto{andrewr@math.ubc.ca}}

\begin{abstract}
We find the exact solution of three interacting friendly directed walks on the square lattice in the bulk, modelling a system of homopolymers that can undergo gelation by introducing two distinct interaction parameters that differentiate between the zipping of only two or all three walks. We establish functional equations for the model's corresponding generating function that are subsequently solved exactly by means of the \emph{obstinate kernel method}.
We then proceed to analyse our model, first considering the case where triple-walk interaction effects are ignored, finding that our model exhibits two phases which we classify as free and gelated regions, with the system exhibiting a second-order phase transition. We then analyse the full model where both interaction parameters are incorporated, presenting the full phase diagram and highlighting the additional existence of a first-order gelation boundary.
\end{abstract}

 \submitto{J. Phys. A.: Math.\ Theor.} 
\maketitle
%
\section{Introduction} 
\label{3fw_intro}
To model the phase behaviour of polymer gelation requires the consideration of 
systems of multiple polymers \cite{Elias2005}. The study of two 
polymers with interpolymer interactions and interacting with a surface has 
received much recent attention because of connections to modelling the unzipping 
of DNA. Typically these have been modelled via either self-avoiding or directed 
walk systems on lattices in two and three dimensions with various types of 
contact interactions \cite{essevaz-roulet1997a-a, lubensky2000a-a, 
lubensky2002a-a, orlandini2001a-a, marenduzzo2002a-a, marenduzzo2003a-a, 
marenduzzo2009a-a, owczarek2012exact, tabbara2014exact}. The exact solution of 
directed friendly walkers on the square lattice with such interactions 
\cite{owczarek2012exact, tabbara2014exact} has led to the extension of a key 
combinatorial technique for lattice paths, the obstinate kernel method 
\cite{bousquet2002counting}. 

To extend these integrable models to the problem of polymer gelation we solve a 
system of three polymer strands with contact interactions. We model this system 
by an ensemble of three directed friendly walkers with shared-vertex 
interactions on the square lattice. We introduce two different types of contact 
interactions that differentiate between situations in which two walks share the 
same site or all three walks share the same site.

\par
We begin in \Sref{3fw_model} by constructing our model, first defining the combinatorial class of allowed configurations and subsequently introducing interaction parameters to assign our configurations with corresponding Boltzmann weights. In particular, we incorporate two distinct interaction parameters to differentiate between the zipping of only two or all three walks.
\par
In \Sref{3fw_functional_eqns}, we introduce two further variables that mark the final vertical distances between our three walks for any given configuration. These auxiliary variables, known as \emph{catalytic} variables, are integral to solving our model. We then establish a mapping between our class of allowed triple-walks onto itself which leads to a functional equation for the model's corresponding generating function that incorporates our added parameters.
\par
We then proceed in \Sref{3fw_soln_eql_case} to determine an exact solution to the model's generating function by means of the \emph{obstinate kernel method} \cite{bousquet2002counting}. While the beginnings of \Sref{3fw_soln_eql_case} outline the precise steps undertaken, we briefly mention that this technique consists of generating a finite system of distinct functional equations by applying a set of different transformations to our original relation determined in \Sref{3fw_functional_eqns}. We then subsequently collapse our system to construct a new refined functional equation which provides us (after some further work) with a solution to our generating function.
\par
Equipped with our exact solution of the generating function, we proceed to analyse our model in \Sref{3fw_analysis}. Specifically, \Sref{3fw_sing_analysis_c1} first considers the dominant singularity behaviour of the generating function for the simplified model where we ignore triple contact effects. We find that such a model exhibits a single critical point, arising in two distinct phases of our system --- namely, a \emph{free} and \emph{gelated} phase. In  \Sref{trans:gc1} we determine that the free to gelated phase transition is a second-order, with a finite jump discontinuity in the second-derivative of free energy. 
\par
In \Sref{3fw_sing_analysis_c_full} we extend our analysis to the full symmetric model that incorporates both double and triple interaction effects, specifying the regions of the phases and plotting the phase diagram. While the full phase space is similarly partitioned into two distinct phases, we find the existence of an additional \emph{first-order} phase boundary for relatively low double and high triple interaction Boltzmann weights.
\par
Finally, in \Sref{3fw_analysis_isolated_parameters}, by a simple 
re-parameterisation, we consider the model that purely isolates double and 
triple walker interaction effects. We plot the new phase diagram and use low and 
high-temperature arguments to explain the limiting behaviour of 
our phase boundaries. As one might 
expect, the order of all phase transitions across the entire phase 
space remain unchanged.
%
\section{The model} 
\label{3fw_model}
\begin{figure}[h]
\psfrag{O}{\small{$(0,0)$}}
\centering 
\includegraphics[width=300px]{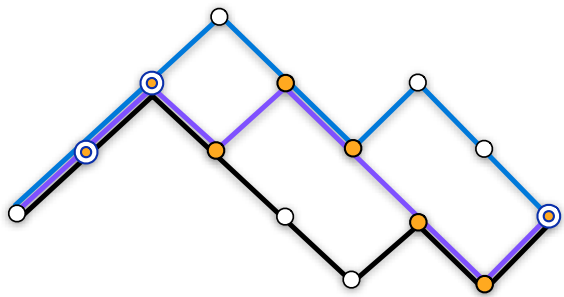} 
\caption[Three friendly interaction walks in the bulk: example of an allowed 
configuration]{An example of an allowed configuration of length 8. Here, we have 
$m_c=11$ double shared contact steps and $m_d=3$ triple shared contact sites. 
Thus, the overall Boltzmann weight for this configuration is $c^{11} d^{3}$ } 
\label{3fw_figure_typical_config} 
\end{figure}

%
Consider three directed walks along the square integer lattice consisting of an equal number of steps. All walks begin at the origin and end at the \emph{same} site. Moreover, walks only can take steps in either the north-east $(1,1)$ or south-east $(1,-1)$ direction. Finally, all three walks may share common steps, however none of the walks are able to cross one another. Such walks are typically referred to as (infinitely) \emph{friendly} walks. Let $\widehat{\allowedpairwalks}$ denote the class of allowed triple walks of \emph{any} length. An example of an allowable configuration is given in \Fref{3fw_figure_typical_config}.
%
\par
For any configuration $\varphi \in \widehat{\allowedpairwalks}$, we assign a 
weight $c$ to the $m_c (\varphi)$ \emph{shared contact sites} between the 
top-to-middle or middle-to-bottom walks respectively. Note, that when all three 
walks share the same site we consider the walk as consisting of \emph{two} 
shared contacts sites with corresponding weight $c^2$ and further the trivial 
triplet of walks of zero length has weight 1. Finally, we assign a weight $d$ to 
the $m_d (\varphi)$ \emph{triple shared contact sites} where all three walks 
coincide, hence contributing a total factor of $c^2 d$ to the overall 
configuration weight. The partition function for our model consisting of $n$ 
paired steps is
%
\begin{equation}
\label{3fw_partition_fn}
Z_n(c,d) =\sum_{\varphi \in \widehat{\allowedpairwalks}, \\ |\varphi| = n}  
c^{\tmcontactsites} d^{\mbcontactsites},
\end{equation}
where $|\varphi|$ denotes the length of the configuration $\varphi$. The reduced free energy $\psi(c,d)$
\begin{equation}
\label{3fw_limiting free_energy}
\psi(c,d) = - \lim_{n \rightarrow \infty}\frac{1}{n} \log{Z_n(c,d)}.
\end{equation}
and generating function $G(c,d;z)$
\begin{equation}
\label{3fw_gen_fn}
G(c,d;z) = \sum_{n=0}^{\infty} Z_n(c,d) z^n
\end{equation}
are defined in the usual manner, where $z$ is conjugate to the length of the configuration.
\par
\begin{figure}[h]
\psfrag{O}{\small{$(0,0)$}}
\centering 
\includegraphics[width=300px]{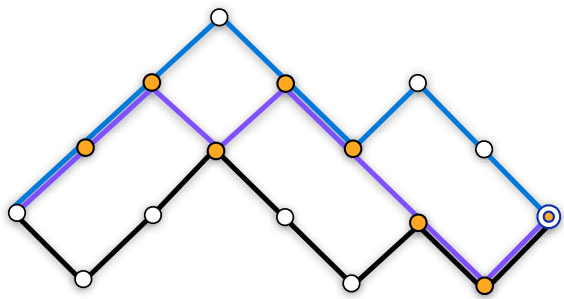} 
\caption{An example of a primitive configuration where all three walks only coincide on the first and last visited sites.} 
\label{3fw_figure_primitve}
\end{figure}

Let $\widehat{\allowedpairwalks}_P$ be the subclass of allowed configurations where \emph{all} three walks share a common site only at the very beginning and end of the configuration. An example of such a configuration is seen in \Fref{3fw_figure_primitve}.
%
We can then define the corresponding \emph{primitive} generating function $P(c;z)$ as
\begin{equation}
P(c;z) = \sum_{\varphi \in \widehat{\allowedpairwalks}_P} z^{|\varphi|}  c^{\tmcontactsites}
\end{equation}
where $z$ is conjugate to the length $|\varphi|$ of a configuration $\varphi \in \widehat{\allowedpairwalks}_P$. Note that we can append any two configurations, $\varphi_1, \varphi_2 \in \widehat{\allowedpairwalks}_P$ together to construct a new configuration $\varphi = \varphi_1 \cdot \varphi_2$ that now lies in our original class $\widehat{\allowedpairwalks}$. More generally any $\varphi \in \widehat{\allowedpairwalks}$ can be uniquely decomposed into a sequence of primitive walks appended to each other as highlighted in \Fref{3fw_figure_primitve}.
%
\begin{figure}[h]
\psfrag{O}{\small{$(0,0)$}}
\centering 
\includegraphics[width=300px]{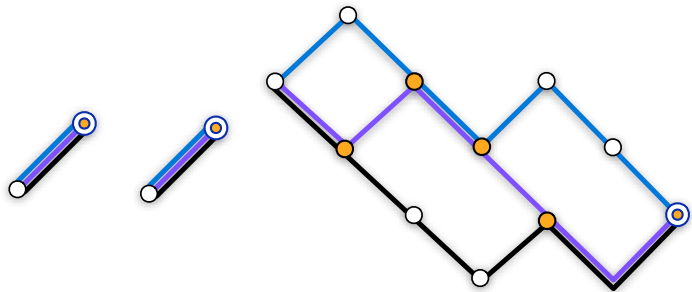} 
\caption[Decomposing an allowed configuration into its primitive components]{Decomposing the configuration seen in \Fref{3fw_figure_typical_config} into its primitive components} \label{3fw_figure_prim_decomp}
\end{figure}
Using the symbolic enumeration formalism described by Flajolet and Sedgewick 
\cite{flajolet2009analytic} we have
\begin{equation}
\widehat{\allowedpairwalks} = SEQ \left({\widehat{\allowedpairwalks}_P} \right) = \left \{ \trivialwalk \right \} + \widehat{\allowedpairwalks}_P + \left \{ \widehat{\allowedpairwalks}_P \times \widehat{\allowedpairwalks}_P \right \} + \left \{ \widehat{\allowedpairwalks}_P \times \widehat{\allowedpairwalks}_P \times \widehat{\allowedpairwalks}_P \right \} + \dots
\end{equation}
where $\left \{ \trivialwalk \right \}$ denotes the trivial configuration consisting of no steps and
\begin{equation}
\left \{ \prod_{i=1}^N \widehat{\allowedpairwalks}_P \right \} = \left \{ \varphi_1 \cdot \varphi_2 \cdot \dots \varphi_N \right | \varphi_i \in \widehat{\allowedpairwalks}_P \}.
\end{equation}
At the level of generating functions, this construction translates to the 
following equation
\begin{equation}
\eqalign {
G(c,d;z) 
&= 1 + d P(c;z) + d^2 {P(c;z)}^2 + \dots \\
&= \frac{1}{1 - dP(c;z)}.
}
\end{equation}
Letting $d=1$ we can express our primitive generating function in terms of $G(c,1;z)$
\begin{equation}
P(c;z) = \frac{G(c,1;z) - 1}{G(c,1;z)},
\end{equation}
which further gives us the relation
\begin{equation}
\label{3fw_full_model_decomp}
G(c,d;z) = \frac{G(c,1;z)}{d \left [1-G(c,1;z) \right ] + G(c,1;z)}.
\end{equation}
Hence to solve for our full model it suffices to solve for the model that ignores triple shared contact effects with corresponding generating function $G(c,1;z)$ which we will indeed proceed to do in \Sref{3fw_functional_eqns} and \Sref{3fw_soln_eql_case}.
%
%
\section{Constructing the functional equations} 
\label{3fw_functional_eqns}
We can establish a functional equation for $G(c,1;z)$ by considering the effect of appending a triplet of steps to the end of any given configuration $\varphi \in \widehat{\allowedpairwalks}$. To begin, we define $\allowedpairwalks(i,j)$ to be the class of triple walks that consists of configurations with final top to middle walk distance $i$ and middle to bottom distance $j$, that still obey friendly constraints. We define our larger combinatorial class $\allowedpairwalks \left(0^{+},  0^{+} \right)$ as
\begin{equation}
\allowedpairwalksfull \equiv \bigcup_{i \geq 0, j \geq 0} \allowedpairwalks(i,j).
\end{equation}
Note that our original class of walks in our model $\widehat{\allowedpairwalks} \equiv \allowedpairwalks(0,0)$. Equipped with our larger combinatorial class we can introduce its corresponding generating function $F(c;z)$ that encodes information about the number of steps and shared contacts for each configuration $\varphi \in \allowedpairwalksfull$. However, determining whether appending a triple-step onto a given configuration $\varphi$ results in a new and \emph{allowable} configuration (i.e. $\varphi$ remains in $\allowedpairwalksfull$) further requires knowledge of the \emph{final} step distances between the three walks. Hence, solely for the purpose of establishing our functional equation for $F(c;z)$, we additionally introduce two catalytic variables $r$ and $s$ to construct the expanded generating function $F(r,s,c;z)$ where
\begin{equation}
\label{3fw_expanded_gen_fn}
F(r,s,c;z) \equiv F(r,s) = \sum_{\varphi \in \allowedpairwalksfull } z^{|\varphi|}  r^{h(\varphi)/2} s^{f(\varphi)/2}  c^{\tmcontactsites}
\end{equation}
and again $z$ is conjugate to the length $|\varphi|$ of a configuration $\varphi \in \allowedpairwalksfull$, $r$ and $s$ are conjugate to \emph{half} the distance $h(\varphi)$ and $f(\varphi)$ between the final vertices of the top to middle and middle to bottom walks respectively. For each $\varphi \in \allowedpairwalksfull$, powers of $r$ and $s$ in $F(r,s)$ track the final step distances between the three walks. Due to the allowed step directions, both $h(\varphi)$ and $f(\varphi)$ must always be even, ensuring that $F(r,s)$ contains only integer powers of $r$ and $s$. Thus, we consider $F(r,s)$ as an element of $\mathbb{Z}[r,s,c][[z]]$: the ring of formal power series in $z$ with coefficients in $\mathbb{Z}[r,s,c]$.
\par
We aim to solve $F(0,0,c;z) \equiv G(c,1;z)$ by establishing a functional equation for $F(r,s)$. Specifically, we construct a suitable mapping from $\allowedpairwalksfull$ onto itself by considering the effect of appending an \emph{allowable} triple-step onto a configuration, translating this map into its action on the generating function. At the end of any given walk we can append a step $(1, \pm 1)$. Hence, for a triplet of walks, there are a total of eight possible combinations of triple steps that can be appended onto a configuration. Let $\mathcal{S}$ be the set of allowable steps, with

{\small
\begin{equation}
\fl
\label{3fw_step_set}
\eqalign {
\mathcal{S} &= \{ (1,1,1), (-1,1,1), (1,-1,1), (1,1,-1), (-1,-1,1), (-1,1,-1), (1,-1,-1), (-1,-1,-1) \}
}
\end{equation}
}
that alter the corresponding configuration weight by factors of $z$, $\frac{z}{r}$, $\frac{zr}{s}$, $zs$, $\frac{z}{s}$, $\frac{zs}{r}$, $zr$ and $z$ respectively. Note, in \eref{3fw_step_set} we have used the shorthand $(x,y,z)$ to denote the triple-step $\{ (1,x), (1,y), (1,z)\}$. 
\par
\begin{figure}
\psfrag{A}{$\mathcal{A}$}
\psfrag{B}{$a$}
\psfrag{C}{$\mathcal{C}$}
\psfrag{D}{$c$}
\centering \subfloat[]{\label{3fw:schematic:uuu}
\includegraphics[width=125px]{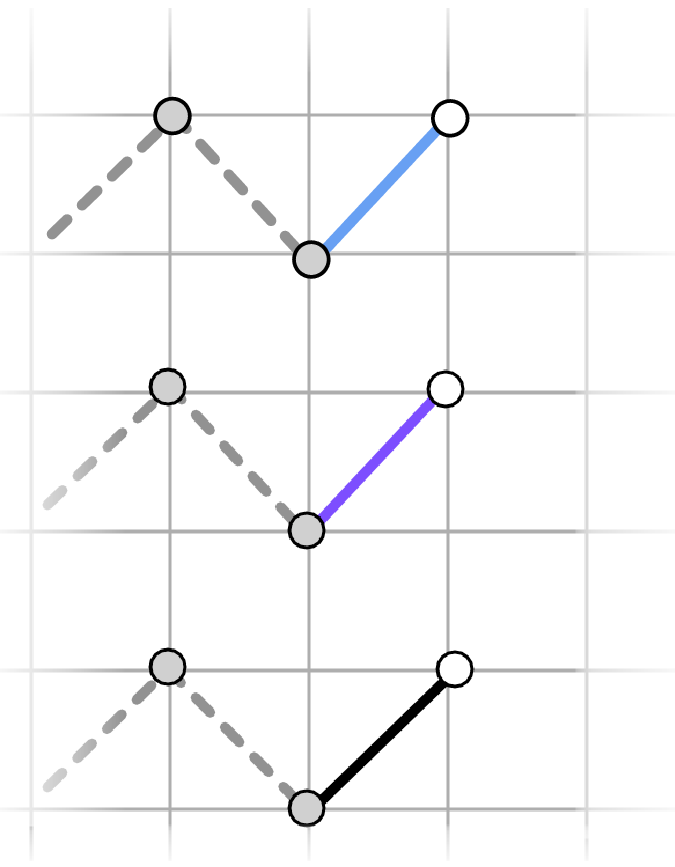}} 
\subfloat[]{\label{3fw:schematic:duu}
\includegraphics[width=125px]{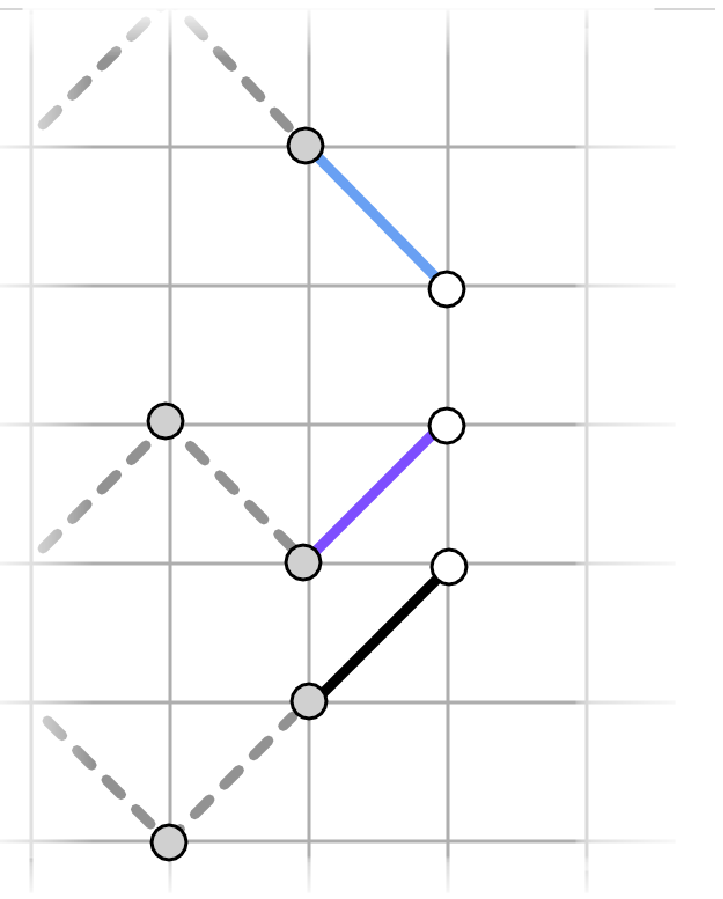}} 
\subfloat[]{\label{3fw:schematic:udu}
\includegraphics[width=125px]{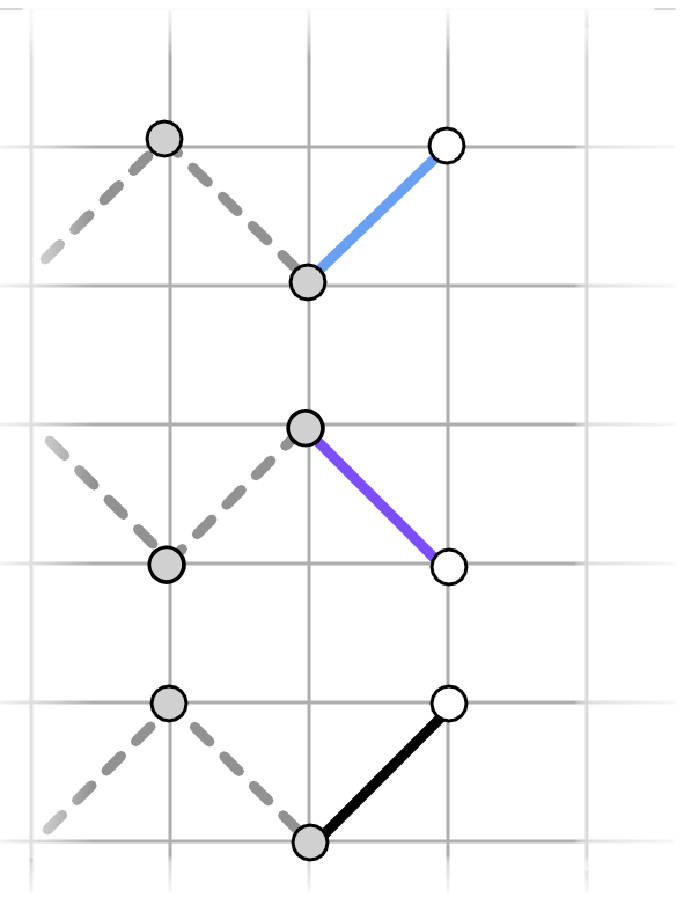}} \\
\subfloat[]{\label{3fw:schematic:uud}
\includegraphics[width=125px]{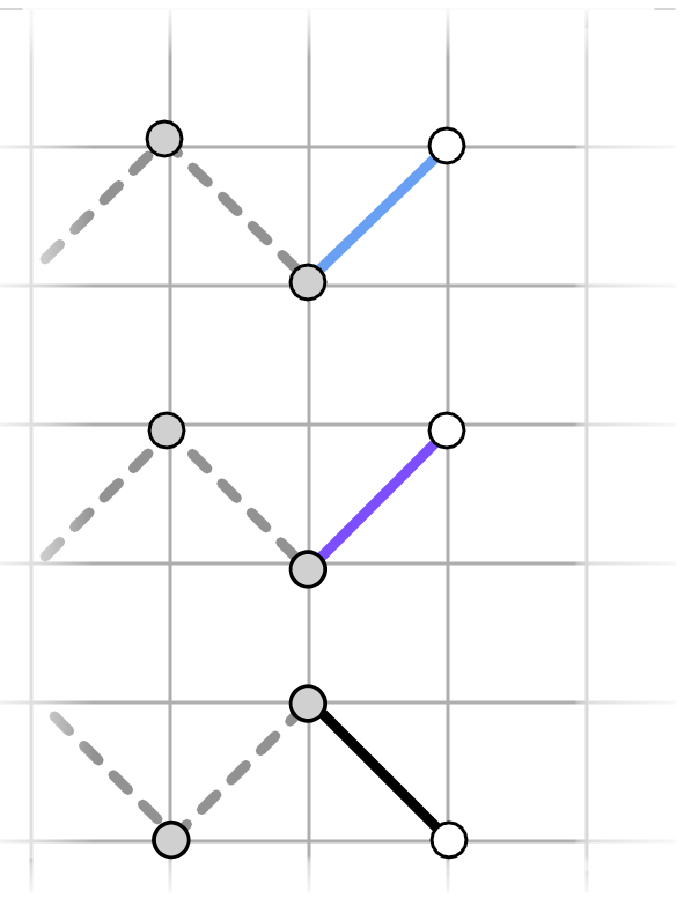}}
\subfloat[]{\label{3fw:schematic:ddu}
\includegraphics[width=125px]{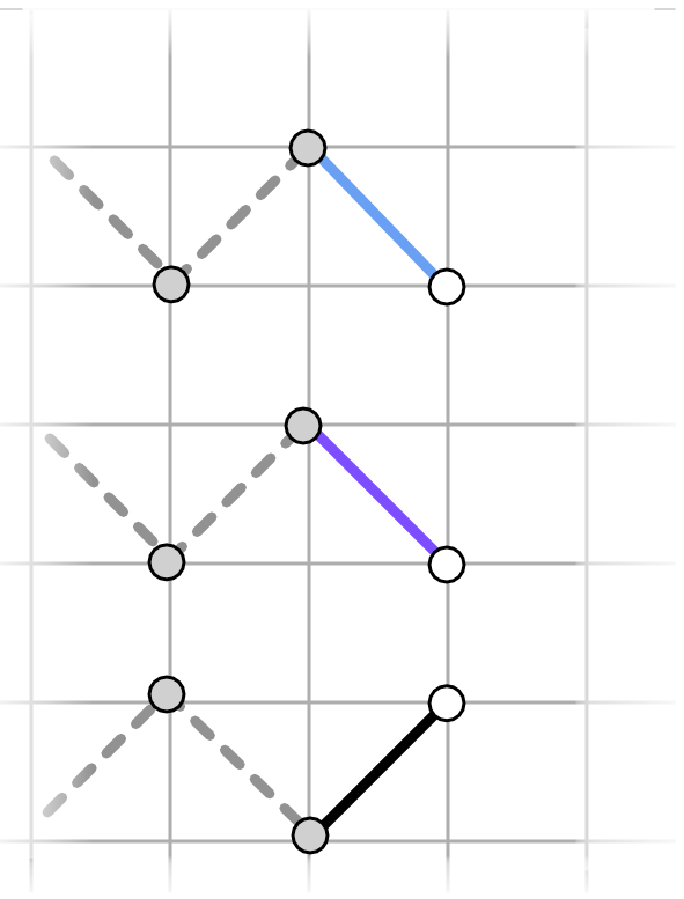}}
\subfloat[]{\label{3fw:schematic:dud}
\includegraphics[width=125px]{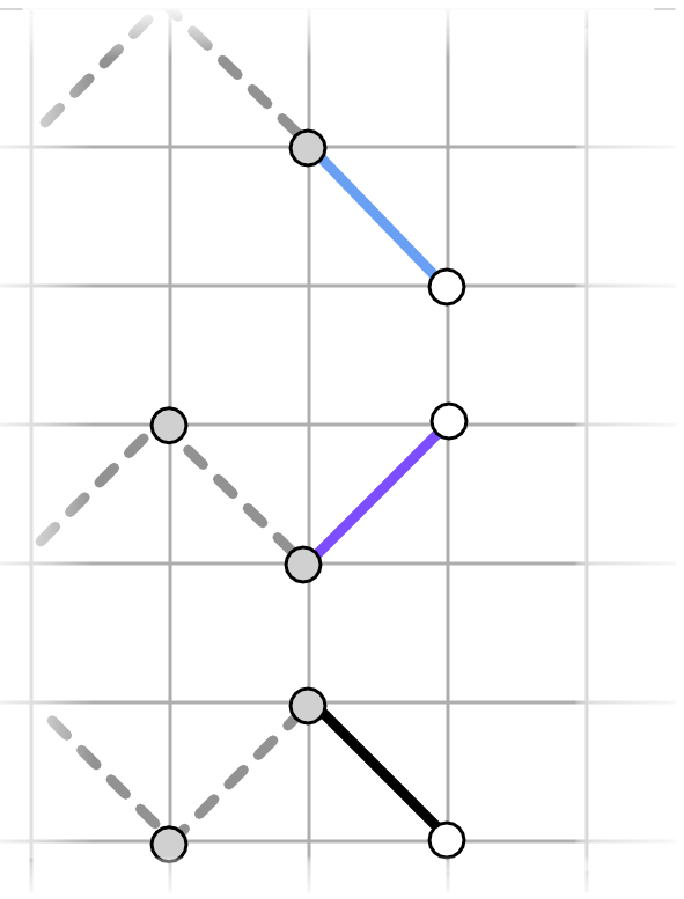}} \\
\subfloat[]{\label{3fw:schematic:udd}
\includegraphics[width=125px]{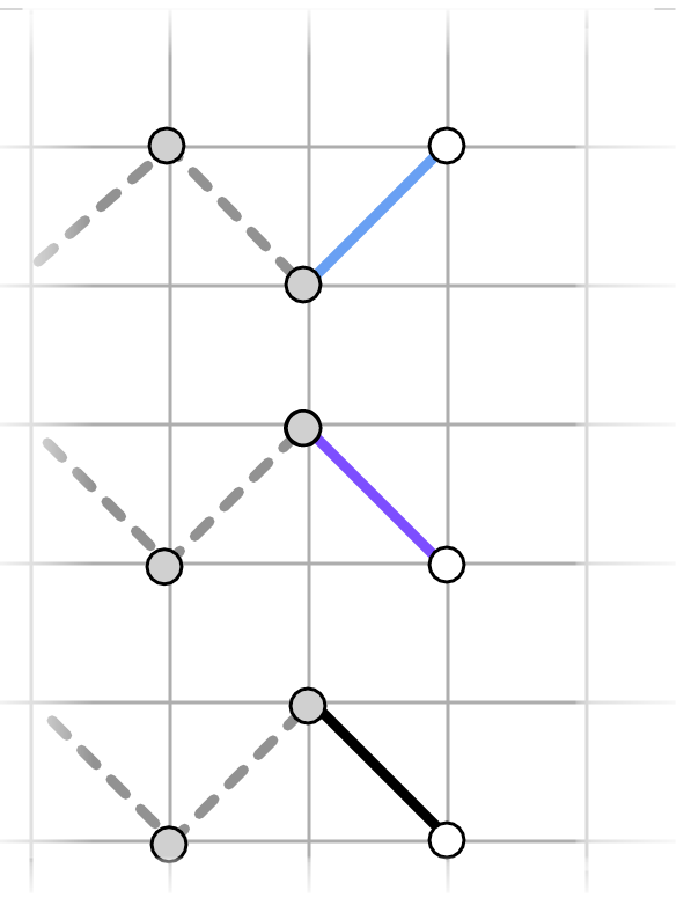}}
\subfloat[]{\label{3fw:schematic:ddd}
\includegraphics[width=125px]{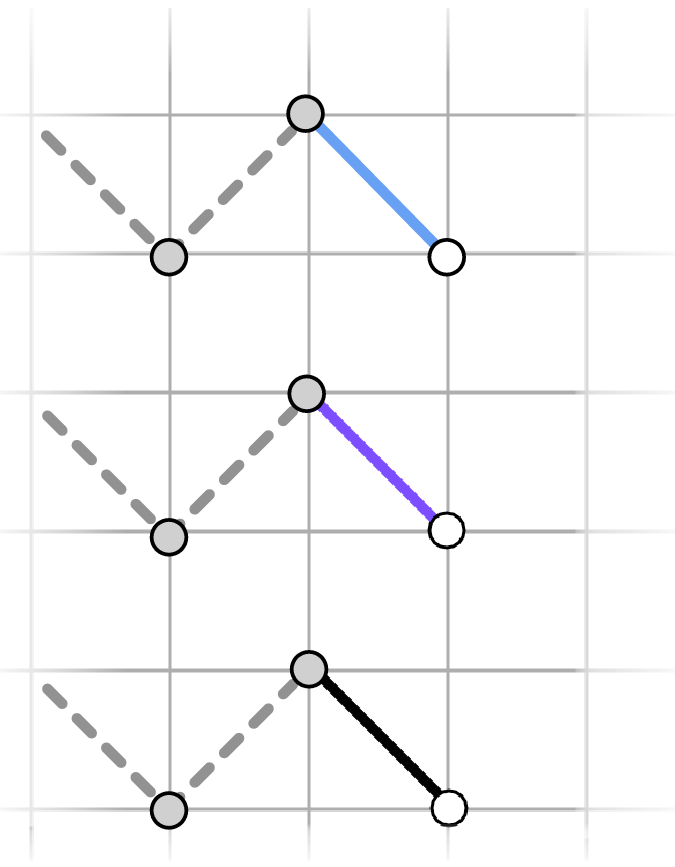}}
\caption{The eight possible ways of appending a triplet of steps to an allowed configuration that results in no new shared contacts.
} 
\label{3fw_schematic_isotropic}
\end{figure}
\begin{figure}
\psfrag{A}{$\mathcal{A}$}
\psfrag{B}{$a$}
\psfrag{C}{$\mathcal{C}$}
\psfrag{D}{$c$}
\centering \subfloat[]{\label{3fw:schematic:uuu:triple}
\includegraphics[width=110px]{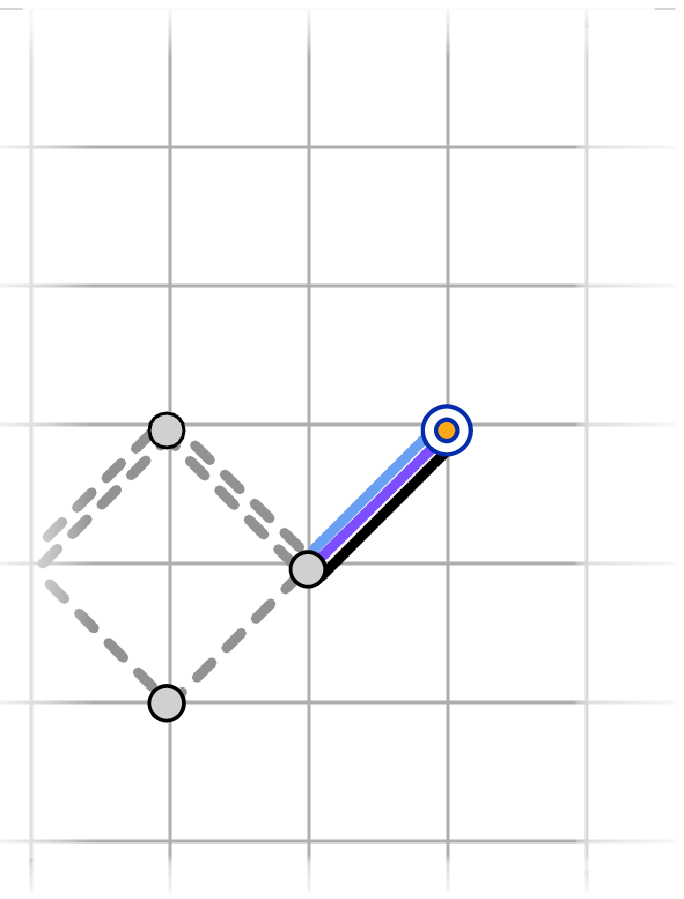}} 
\subfloat[]{\label{3fw:schematic:ddd:triple}
\includegraphics[width=110px]{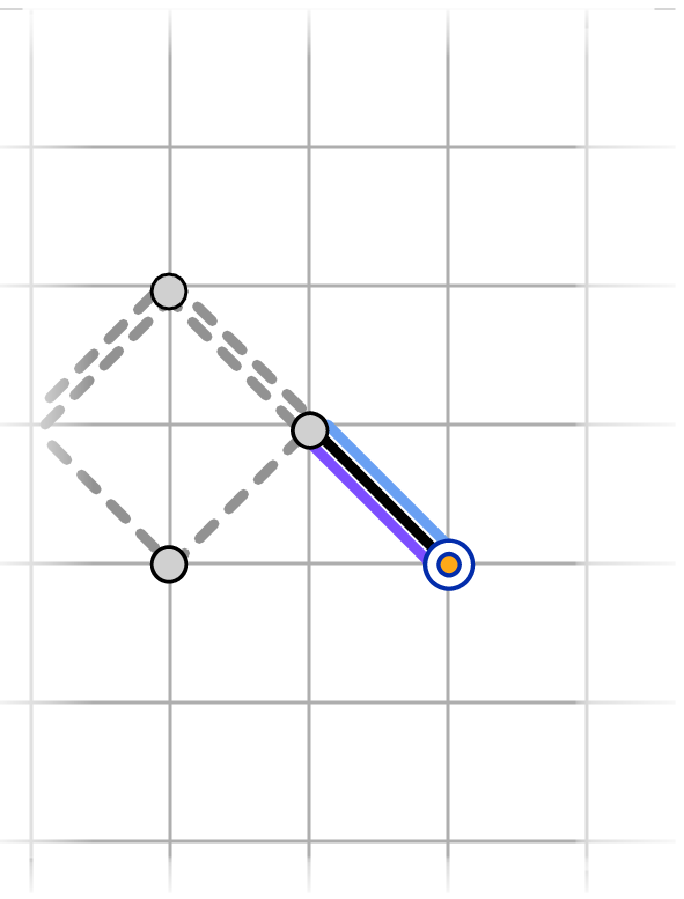}} \\
\subfloat[]{\label{3fw:schematic:ddu:triple}
\includegraphics[width=110px]{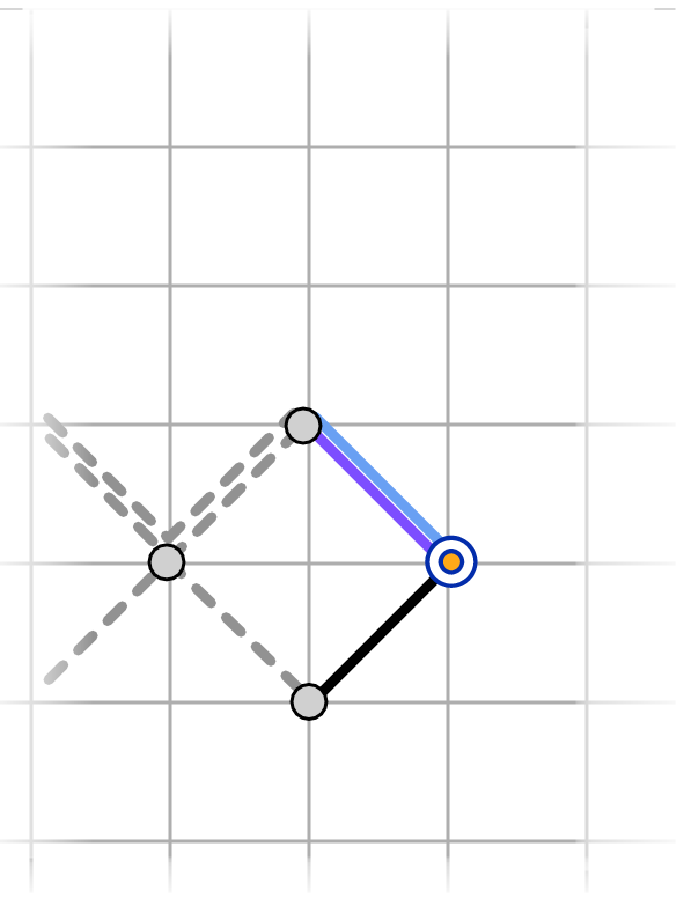}}
\subfloat[]{\label{3fw:schematic:duu:triple}
\includegraphics[width=110px]{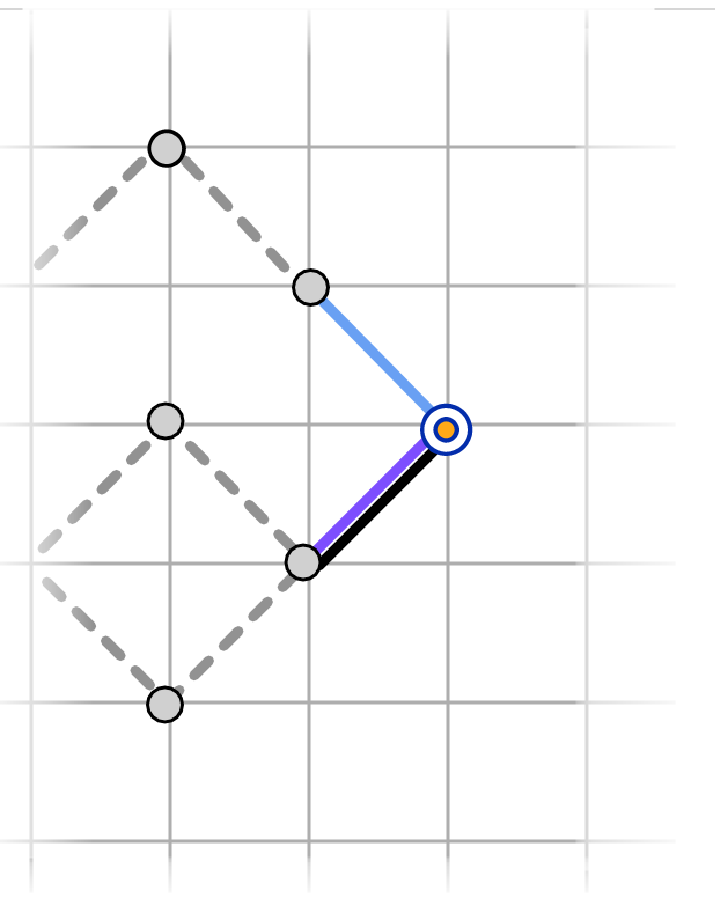}}
\caption{The four possible ways of appending a triplet of steps to an allowed configuration that results in all walks visiting the same site.
} 
\label{3fw_schematic_triple}
\end{figure}
\begin{figure}[h!]
\psfrag{A}{$\mathcal{A}$}
\psfrag{B}{$a$}
\psfrag{C}{$\mathcal{C}$}
\psfrag{D}{$c$}
\centering 
\subfloat[]{\label{3fw:schematic:uud:double}
\includegraphics[width=110px]{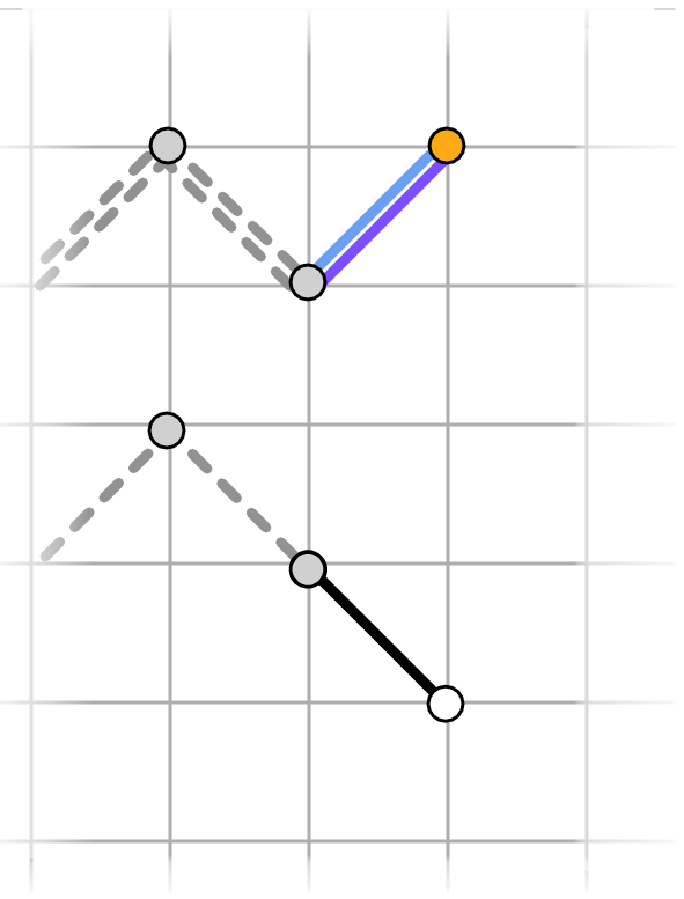}}
\subfloat[]{\label{3fw:schematic:uuu:double}
\includegraphics[width=110px]{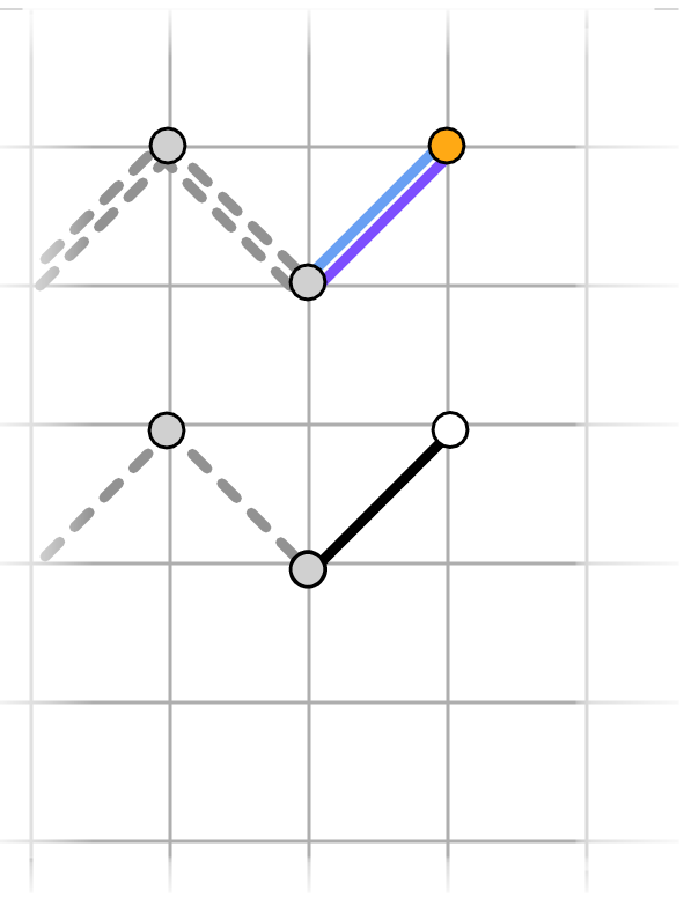}}
\subfloat[]{\label{3fw:schematic:ddd:double2}
\includegraphics[width=110px]{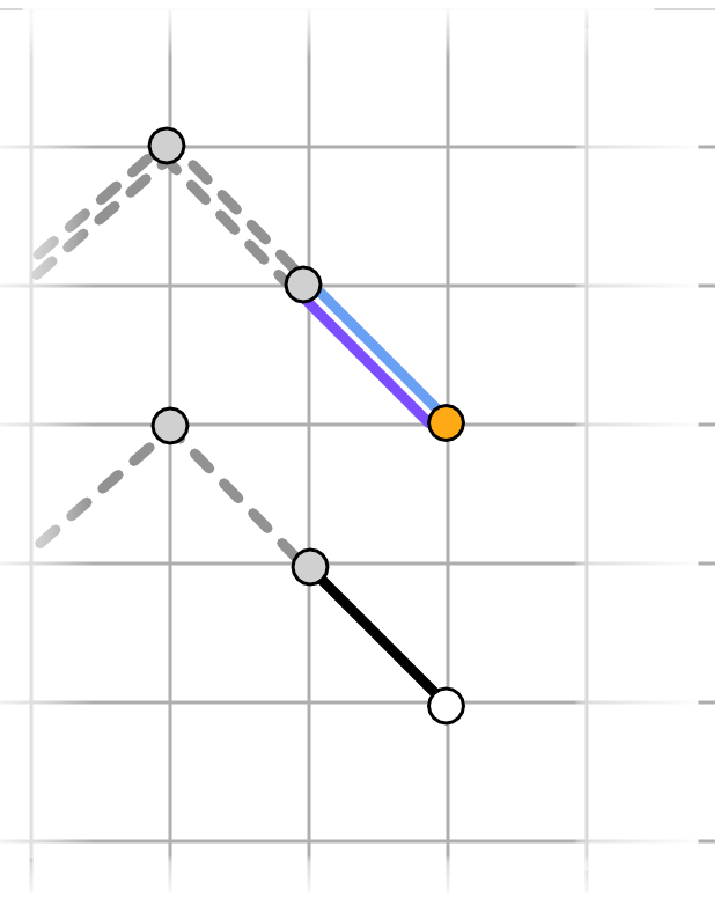}}
\subfloat[]{\label{3fw:schematic:ddu:double2}
\includegraphics[width=110px]{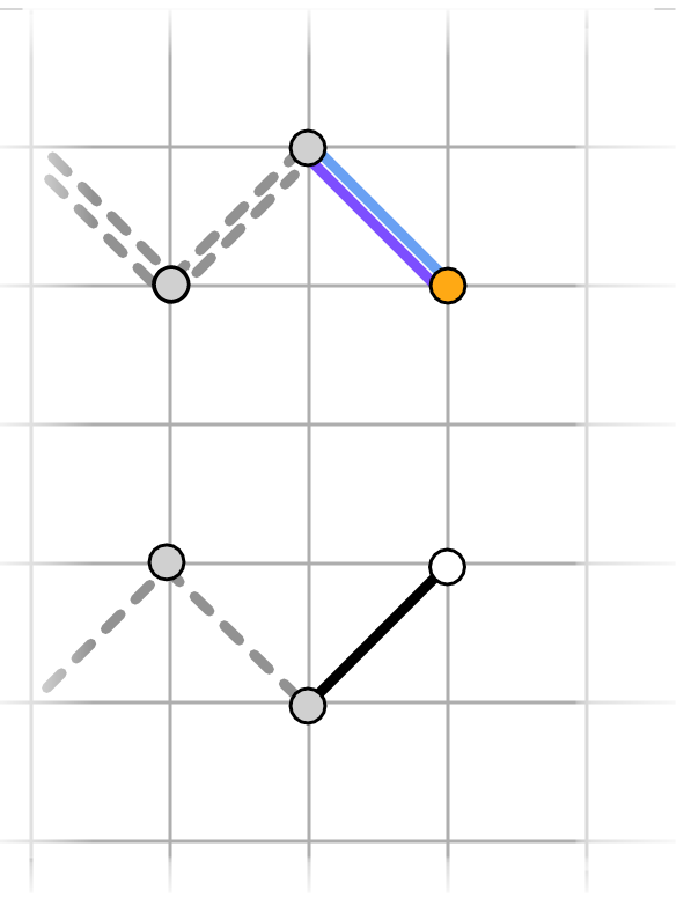}} \\
\subfloat[]{\label{3fw:schematic:ddu:double}
\includegraphics[width=110px]{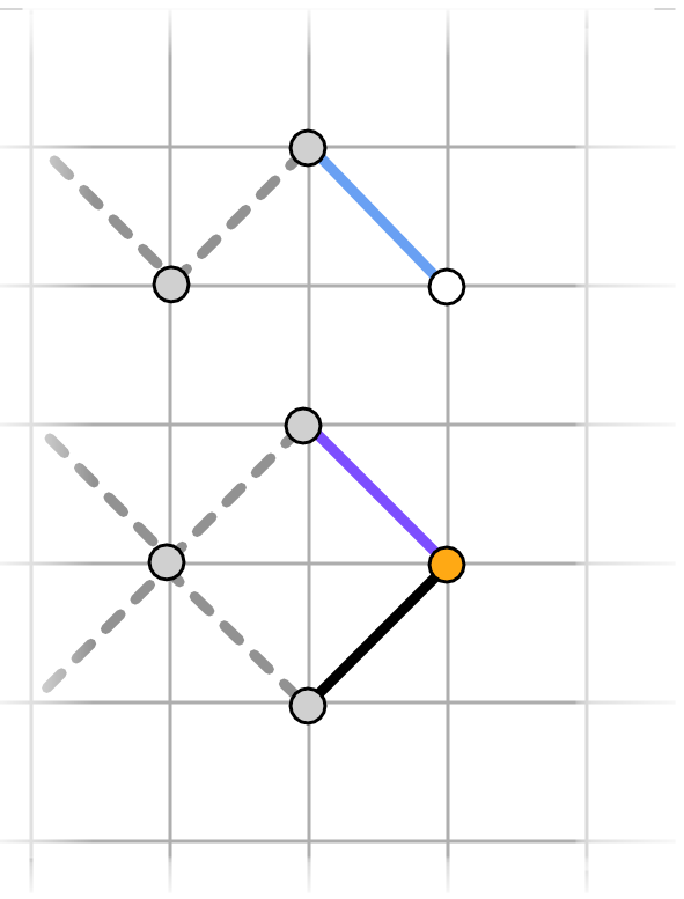}}
\subfloat[]{\label{3fw:schematic:duu:double}
\includegraphics[width=110px]{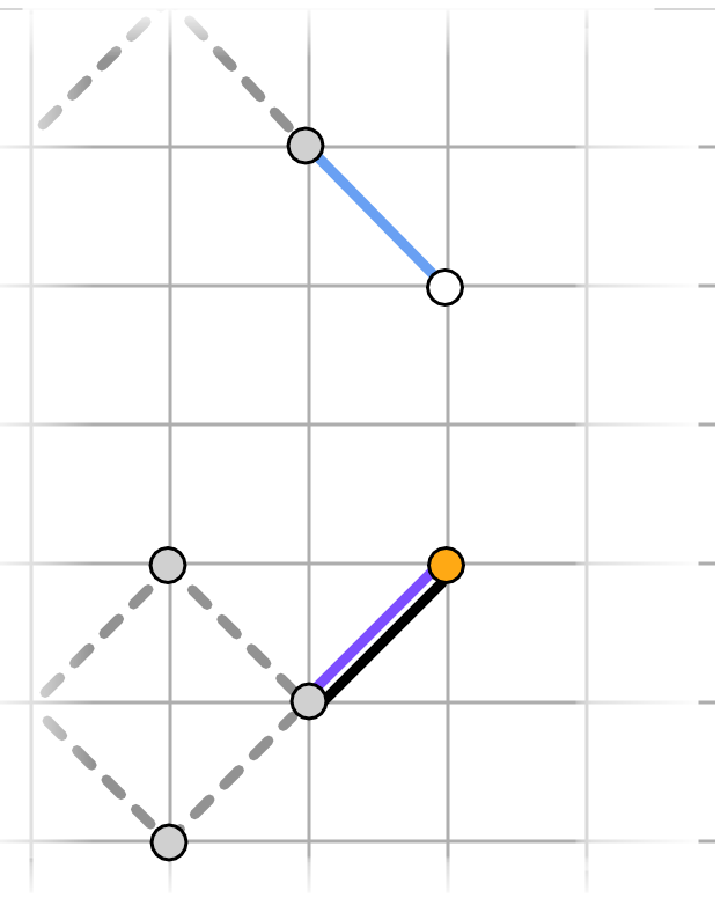}}
\subfloat[]{\label{3fw:schematic:uuu:double2} 
\includegraphics[width=110px]{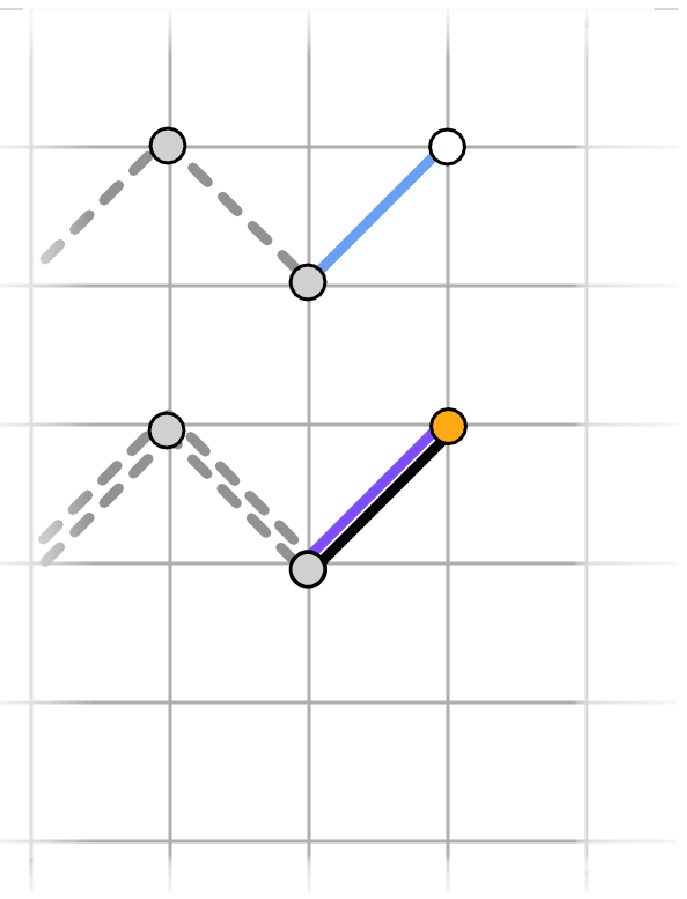}} 
\subfloat[]{\label{3fw:schematic:udd:double}
\includegraphics[width=110px]{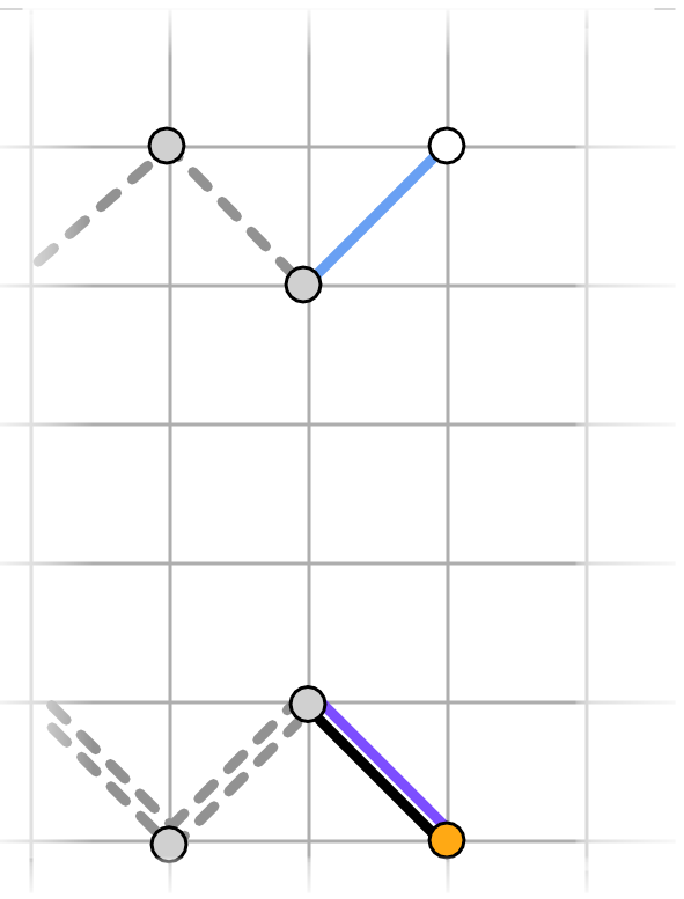}} \\
\subfloat[]{\label{3fw:schematic:ddd:double}
\includegraphics[width=110px]{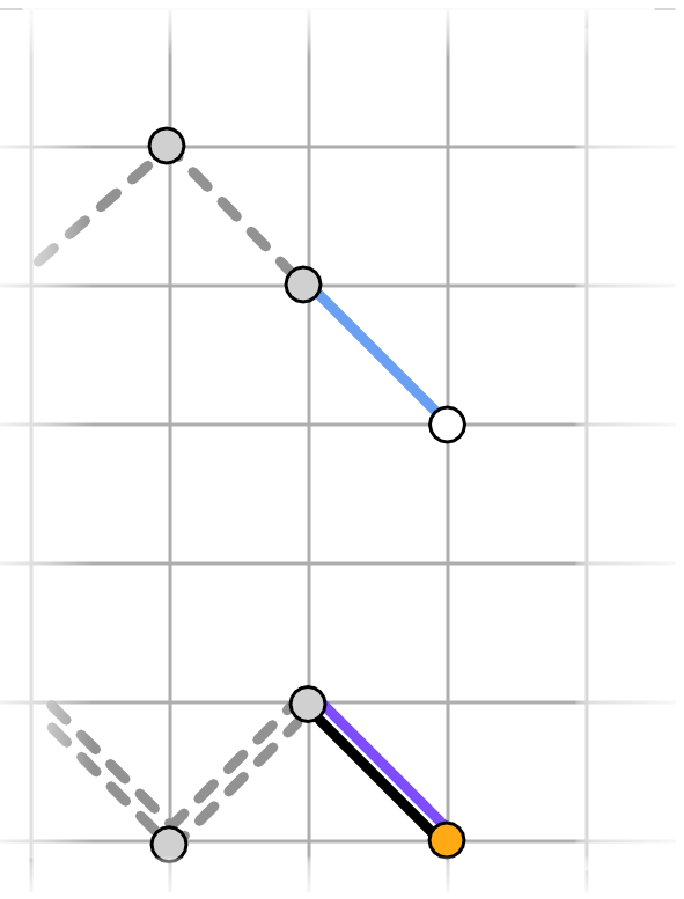}} 
\subfloat[]{\label{3fw:schematic:dud:double}
\includegraphics[width=110px]{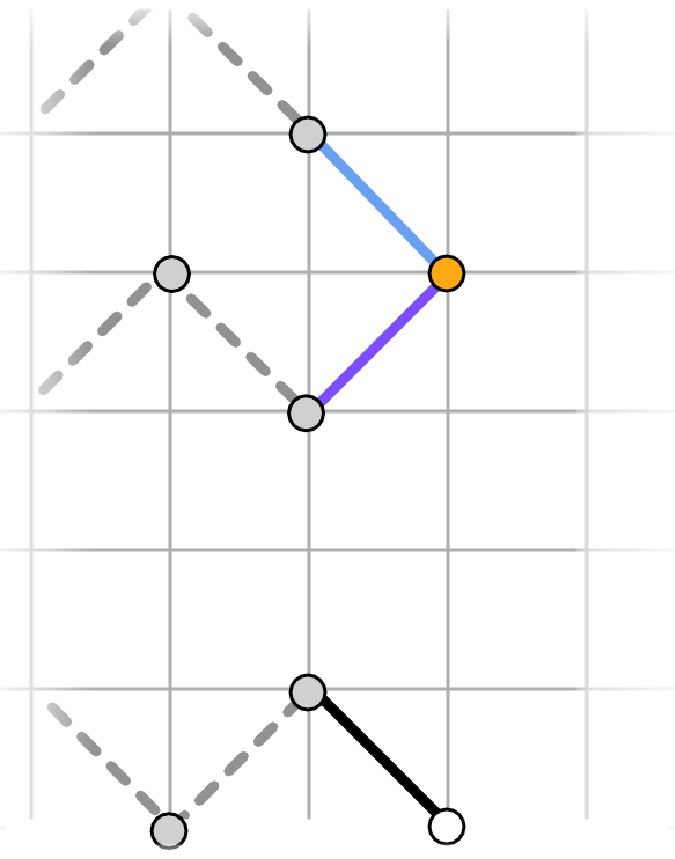}}
\subfloat[]{\label{3fw:schematic:duu:double2}
\includegraphics[width=110px]{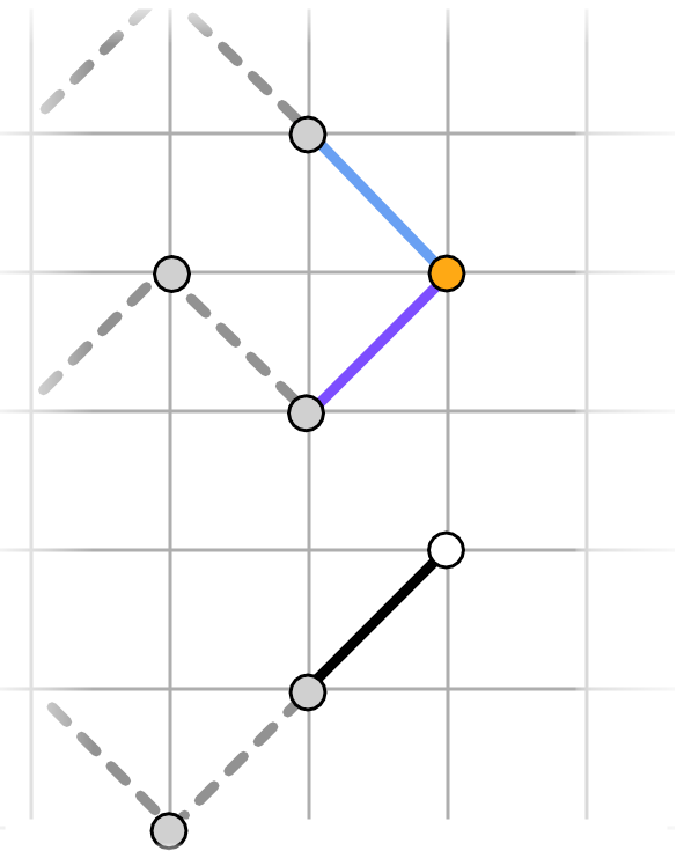}}
\subfloat[]{\label{3fw:schematic:udu:double}
\includegraphics[width=110px]{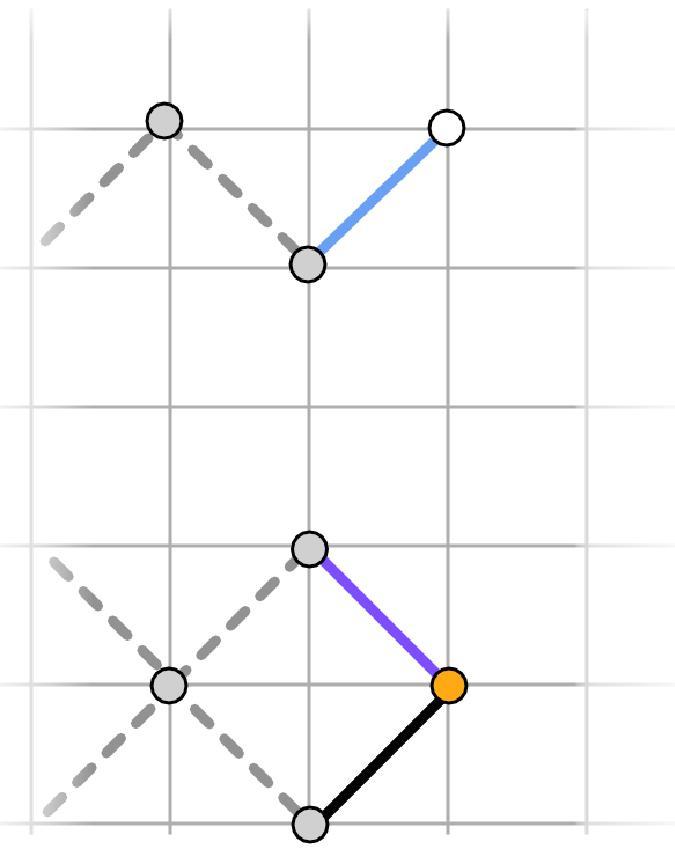}}
\caption{The twelve possible ways of appending a triplet of steps to an allowed configuration where only two of the three walks visit the same site.
} 
\label{figure:3fw:double_contact}
\end{figure}

Given the non-crossing constraint between the three walks, not all eight combinations of appended steps will necessarily result in allowable configurations and furthermore shared contact interaction effects also need to be considered when attaching new steps. Thus, we identify 22 distinct cases that capture all possible changes in weight that can arise from appending a triplet of steps as seen in \Fref{3fw_schematic_isotropic}, \Fref{3fw_schematic_triple} and \Fref{figure:3fw:double_contact}; allowing us to construct a functional equation for $F(r,s)$, highlighting the underlying decomposition for $\allowedpairwalksfull$. 
We denote $\{ \trivialwalk \}$ as the trivial zero-length configuration and introduce the following shorthand notation

{\normalsize
\begin{equation}
\fl
\allowedpairwalks \left(n^{+}, j\right) \equiv \bigcup_{i\geq n} \allowedpairwalks \left(i, j \right), \quad \allowedpairwalks \left(i, m^{+} \right)  \equiv \bigcup_{j \geq m} \allowedpairwalks \left(i, j \right), \quad \allowedpairwalks \left(n^{+}, m^{+}\right) = \bigcup_{i\geq n, j \geq m} \allowedpairwalks \left(i, j \right),
\end{equation}
}
while 
\begin{equation}
\{\sigma \} \cdot \allowedpairwalks \left(i, j \right),
\end{equation}
represents the class of configurations formed by appending the triple-step 
$\sigma \in \mathcal{S}$ to the \emph{end} of each triplet of walks $\varphi 
\in \allowedpairwalks \left(i, j \right)$. We can build up our functional 
equation $F(r,s,c;z)$ by firstly establishing a relation for the non-interacting 
case $F(r,s,1;z)$ and subsequently incorporating the effects of shared contacts. 
To do this we consider the effect of appending a triple-step onto a given 
configuration, making sure to eliminate newly formed walks that are no longer 
part of our allowable class $\allowedpairwalksfull$. For $F(r,s,1;z)$ we find

{\normalsize
\begin{equation}
\label{3fw_main_fn_eqn_isotropic}
\fl \eqalign{
F(r,s,1,1;z) =  \qquad &\allowedpairwalksfull= \\
\nonumber 1 \qquad  &\{ \trivialwalk \} \\
\nonumber + z  F(r,s) \qquad  & \bigcup\{ (1,1,1) \} \cdot \allowedpairwalksfull, \quad 
\mbox{\Fref{3fw:schematic:uuu}} \\
\nonumber + \frac{z}{r} \left ( F(r,s) - [r^0] F(r,s) \right ) \qquad  & \bigcup\{ (-1,1,1) \} \cdot \allowedpairwalks \left(1^{+}, 0^{+} \right), \quad 
\mbox{\Fref{3fw:schematic:duu}} \\
\nonumber + \frac{zr}{s} \left ( F(r,s) - [s^0] F(r,s) \right ) \qquad  & 
\bigcup\{ (1,-1,1) \} \cdot \allowedpairwalks \left(0^{+}, 1^{+} \right), \quad 
\mbox{\Fref{3fw:schematic:udu}}  \\
\nonumber + zs F(r,s) \qquad  & \bigcup\{ (1,1,-1) \} \cdot \allowedpairwalksfull, 
\quad \mbox{\Fref{3fw:schematic:uud}} \\
\nonumber + \frac{z}{s} \left ( F(r,s) - [s^0] F(r,s) \right ) \qquad  & 
\bigcup\{ (-1,-1,1) \} \cdot \allowedpairwalks \left(0^{+}, 1^{+} \right), \quad 
\mbox{\Fref{3fw:schematic:ddu}}  \\
\nonumber + \frac{zs}{r} \left ( F(r,s) - [r^0] F(r,s) \right ) \qquad  &  
\bigcup\{(-1,1,-1) \} \cdot \allowedpairwalks \left(1^{+}, 0^{+} \right), \quad 
\mbox{\Fref{3fw:schematic:dud}}  \\
\nonumber + zr F(r,s) \qquad  &  \bigcup\{(1,-1,-1) \} \cdot 
\allowedpairwalksfull, \quad \mbox{\Fref{3fw:schematic:udd}}  \\
\nonumber + z F(r,s) \qquad  &  \bigcup\{(-1,-1,-1) \} \cdot 
\allowedpairwalksfull, \quad \mbox{\Fref{3fw:schematic:ddd}},
}
\end{equation}
}
where $[r^j] F(r,s)$,$[s^k] F(r,s)$ and in general $[r^j s^k] F(r,s)$ denote the 
coefficients of $r^j$, $s^k$ and $r^j s^k$ in the generating function $F(r,s)$ 
respectively. Note, that since the coefficients of $F(r,s)$ are polynomials 
in $r,s$ we have
\begin{equation}
\eqalign{
[r^0] F(r,s) = F(0,s)  \\
[s^0] F(r,s) = F(r,0), \\
[r^0 s^0] F(r,s) = F(0,0).
}
\end{equation}
Next, we add shared contact site effects to \eref{3fw_main_fn_eqn_isotropic} to get a functional equation for $F(r,s,c;z)$, with
{\small
\begin{equation}
\label{3fw_main_fn_eqn_shared_contacts}
\fl \eqalign{
F(r,s,c,1;z) = \mbox{RHS of \eref{3fw_main_fn_eqn_isotropic}}  \qquad & \\
\nonumber + z (c^2-1) F(0,0) 
\qquad  & \bigcup\{ (1,1,1) \} \cdot \alw{0}{0}, \quad 
\mbox{\Fref{3fw:schematic:uuu:triple}} \\
\nonumber + z (c^2 -1) F(0,0) 
\qquad  & \bigcup\{ (-1,-1,-1) \} \cdot \alw{0}{0}, \quad 
\mbox{\Fref{3fw:schematic:ddd:triple}} \\
\nonumber + z(c^2-1) [s^1] F(0,s)
\qquad  & \bigcup\{ (-1,-1,1) \} \cdot \alw{0}{2}, \quad 
\mbox{\Fref{3fw:schematic:ddu:triple}} \\
\nonumber + z(c^2-1) [r^1] F(r,0)
\qquad  & \bigcup\{ (-1,1,1) \} \cdot \alw{2}{0}, \quad 
\mbox{\Fref{3fw:schematic:duu:triple}} \\
\nonumber + z (c-1) F(0,s) 
\qquad  & \bigcup\{ (1,1,-1) \} \cdot \alw{0}{0^+}, \quad 
\mbox{\Fref{3fw:schematic:uud:double}} \\
\nonumber + z (c-1) \left ( F(0,s) - F(0,0) \right )
\qquad  & \bigcup\{ (1,1,1) \} \cdot \alw{0}{2^+}, \quad 
\mbox{\Fref{3fw:schematic:uuu:double}} \\
\nonumber + z (c-1)\ \left ( F(0,s) - F(0,0) \right) \qquad & \bigcup \{ -1, -1, -1 \} \cdot \alw{0}{2^+}, \quad \mbox{\Fref{3fw:schematic:ddd:double2}} \\
\nonumber + \frac{z}{s} (c-1)\ \left ( F(0,s) - F(0,0) - s[s^1] F(0,s) \right) \qquad & \bigcup \{ -1, -1, 1 \} \cdot \alw{0}{4^+}, \quad \mbox{\Fref{3fw:schematic:ddu:double2}} \\
\nonumber + z (c-1) \left ( [s^1]F(r,s) -  [s^1] F(0,s) \right) 
\qquad  & \bigcup\{ (-1,-1,1) \} \cdot \alw{2^+}{2}, \quad 
\mbox{\Fref{3fw:schematic:ddu:double}} \\
\nonumber + \frac{z}{r} (c-1) \left (F(r,0) - F(0,0) - r[r^1]F(r,0) \right)
\qquad  & \bigcup\{ (-1,1,1) \} \cdot \alw{4^+}{0}, \quad 
\mbox{\Fref{3fw:schematic:duu:double}} \\
\nonumber + zr (c-1) \left (F(r,0) - F(0,0) \right) \qquad  & \bigcup\{ (1,1,1) \} \cdot \alw{2^+}{0}, \quad 
\mbox{\Fref{3fw:schematic:uuu:double2}} \\
\nonumber + zr(c-1) F(r,0) 
\qquad  & \bigcup\{ (1,-1,-1) \} \cdot \alw{0^+}{0}, \quad 
\mbox{\Fref{3fw:schematic:udd:double}} \\
\nonumber + z (c-1) \left ( F(r,0) - F(0,0)\right )
\qquad  & \bigcup\{ (-1,-1,-1) \} \cdot \alw{2^+}{0}, \quad 
\mbox{\Fref{3fw:schematic:ddd:double}} \\
\nonumber + zs(c-1) [r^1] F(r,s) 
\qquad  & \bigcup\{ (-1,1,-1) \} \cdot \alw{2}{0^+}, \quad 
\mbox{\Fref{3fw:schematic:dud:double}} \\
\nonumber + z(c-1) \left ([r^1] F(r,s) - [r^1] F(r,0) \right )
\qquad  & \bigcup\{ (-1,1,1) \} \cdot \alw{2}{2^+}, \quad 
\mbox{\Fref{3fw:schematic:duu:double2}} \\
\nonumber + zr(c-1)[s^1] F(r,s)
\qquad  & \bigcup\{ (1,-1,1) \} \cdot \alw{0^+}{2}, \quad 
\mbox{\Fref{3fw:schematic:udu:double}}
}
\end{equation}
}
%
%
%
%
\clearpage
We now further refine \eref{3fw_main_fn_eqn_shared_contacts} by eliminating the 
terms $[s^1]F(0,s), [r^1]F(r,0)$, $[s^1]F(r,s)$ and $[r^1]F(r,s)$. The first 
step in this process is to construct 3 new functional equations for
$F(0,0), F(0,s)$ and $F(r,0)$:
\begin{equation}
\label{3fw_F00_fn_eqn}
\fl
\eqalign {
F(0,0) = \qquad &\alw{0}{0} = \\
\nonumber 1 \qquad  & \{ \trivialwalk \} \\
\nonumber + z c^2 F(0,0)
\qquad  & \bigcup\{ (1,1,1) \} \cdot \alw{0}{0} \quad 
\mbox{\Fref{3fw:schematic:uuu:triple}} \\
\nonumber + z c^2 F(0,0) 
\qquad  & \bigcup\{ (-1,-1,-1) \} \cdot \alw{0}{0} \quad 
\mbox{\Fref{3fw:schematic:ddd:triple}} \\
\nonumber + z c^2 [s^1] F(0,s)
\qquad  & \bigcup\{ (-1,-1,1) \} \cdot \alw{0}{1} \quad 
\mbox{\Fref{3fw:schematic:ddu:triple}} \\
\nonumber + z c^2 [r^1] F(r,0)
\qquad  & \bigcup\{ (-1,1,1) \} \cdot \alw{1}{0} \quad 
\mbox{\Fref{3fw:schematic:duu:triple}}
}
\end{equation}
and
{\small
\begin{equation}
\label{3fw_F0s_fn_eqn}
\fl
\eqalign{
F(0,s) = \qquad &\allowedpairwalks \left(0, 0^{+} \right) = \\
\nonumber 1 \qquad & \{ \trivialwalk \} \\
+ z c^2 F(0,0) 
\nonumber \qquad  & \bigcup\{ (1,1,1) \} \cdot \alw{0}{0},  \quad 
\mbox{\Fref{3fw:schematic:uuu:triple}} \\
+ z c^2 F(0,0) 
\nonumber \qquad  & \bigcup\{ (-1,-1,-1) \} \cdot \alw{0}{0},  \quad 
\mbox{\Fref{3fw:schematic:ddd:triple}} \\
\nonumber + z c^2 [s^1] F(0,s)
\qquad  & \bigcup\{ (-1,-1,1) \} \cdot \alw{0}{1} \quad 
\mbox{\Fref{3fw:schematic:ddu:triple}} \\
\nonumber + z c^2 [r^1] F(r,0)
\qquad  & \bigcup\{ (-1,1,1) \} \cdot \alw{1}{0} \quad 
\mbox{\Fref{3fw:schematic:duu:triple}} \\
+ zs c F(0,s) 
\nonumber \qquad  & \bigcup\{ (1,1,-1) \} \cdot \alw{0}{0^+}, \quad 
\mbox{\Fref{3fw:schematic:uud:double}} \\
+ z c \left ( F(0,s) - F(0,0) \right) 
\nonumber \qquad  & \bigcup\{ (1,1,1) \} \cdot \alw{0}{2^+}, \quad 
\mbox{\Fref{3fw:schematic:uuu:double}} \\
\nonumber + z c\ \left ( F(0,s) - F(0,0) \right) \qquad & \bigcup \{ -1, -1, -1 \} \cdot \alw{0}{2^+}, \quad 
\mbox{\Fref{3fw:schematic:ddd:double2}} \\
\nonumber + \frac{z}{s} (c-1)\ \left ( F(0,s) - F(0,0) - s[s^1] F(0,s) \right) \qquad & \bigcup \{ -1, -1, 1 \} \cdot \alw{0}{4^+}, \quad \mbox{\Fref{3fw:schematic:ddu:double2}} \\
+ zs c \left ( [r^1]F(r,s)\right)
\nonumber \qquad  & \bigcup\{ (-1,1,-1) \} \cdot \alw{2}{0^+}, \quad 
\mbox{\Fref{3fw:schematic:dud:double}} \\
\nonumber + zc\left ([r^1] F(r,s) - [r^1] F(r,0) \right )
\qquad  & \bigcup\{ (-1,1,1) \} \cdot \alw{2}{2^+}, \quad 
\mbox{\Fref{3fw:schematic:duu:double2}}
}
\end{equation}
}
and
{\small
\begin{equation}
\label{3fw_Fr0_fn_eqn}
\fl
\eqalign{
F(r,0) = \qquad &\allowedpairwalks \left(0^{+}, 0 \right) = \\
\nonumber 1 \qquad & \{ \trivialwalk \} \\
+ z c^2 F(0,0) 
\nonumber \qquad  & \bigcup\{ (1,1,1) \} \cdot \alw{0}{0}, \quad 
\mbox{\Fref{3fw:schematic:uuu:triple}} \\
+ z c^2 F(0,0) 
\nonumber \qquad  & \bigcup\{ (-1,-1,-1) \} \cdot \alw{0}{0}, \quad 
\mbox{\Fref{3fw:schematic:ddd:triple}} \\
\nonumber + z c^2 [s^1] F(0,s)
\qquad  & \bigcup\{ (-1,-1,1) \} \cdot \alw{0}{1} \quad 
\mbox{\Fref{3fw:schematic:ddu:triple}} \\
\nonumber + z c^2 [r^1] F(r,0)
\qquad  & \bigcup\{ (-1,1,1) \} \cdot \alw{1}{0} \quad 
\mbox{\Fref{3fw:schematic:duu:triple}} \\
+ zr c F(r,0) 
\nonumber \qquad  & \bigcup\{ (-1,-1,1) \} \cdot \alw{0^+}{0}, \quad 
\mbox{\Fref{3fw:schematic:ddu:double}} \\
+ z c \left ( F(r,0) - F(0,0) \right) 
\nonumber \qquad  & \bigcup\{ (-1,-1,-1) \} \cdot \alw{2^+}{0}, \quad 
\mbox{\Fref{3fw:schematic:ddd:double}} \\
\nonumber + z c\ \left ( F(r,0) - F(0,0) \right) \qquad & \bigcup \{ 1, 1, 1 \} \cdot \alw{2^+}{0}, \quad 
\mbox{\Fref{3fw:schematic:uuu:double2}} \\
\nonumber + \frac{z}{r} (c-1)\ \left ( F(r,0) - F(0,0) - r[r^1] F(r,0) \right) \qquad & \bigcup \{ 1, 1, -1 \} \cdot \alw{4^+}{0}, \quad \mbox{\Fref{3fw:schematic:uud:double}} \\
+ zr c \left ( [s^1]F(r,s)\right)
\nonumber \qquad  & \bigcup\{ (1,-1,1) \} \cdot \alw{0^+}{2}, \quad 
\mbox{\Fref{3fw:schematic:udu:double}} \\
\nonumber + zc \left ([s^1] F(r,s) - [s^1] F(r,0) \right )
\qquad  & \bigcup\{ (1,-1,-1) \} \cdot \alw{2^+}{2}, \quad 
\mbox{\Fref{3fw:schematic:udd:double}}.
}
\end{equation}
}
Next we solve this system for three of the terms we wish to eliminate (say 
$[s^1]F(0,s), [r^1]F(r,0)$ and $[s^1]F(r,s)$), and then subsitute these 
back into \eref{3fw_main_fn_eqn_shared_contacts}. In so doing, we find that the 
fourth term,  $[r^1]F(r,s)$ , is also eliminated. This results in 
an equation satisfied by $F(r,s)$ that is considerably simpler:%
\begin{equation}
\label{3fw_main_fn_eqn_simplified_v1}
\eqalign {
K(r,s) F(r,s) &= \frac{1}{c^2} - \frac{(r-c r+c z+c s z)}{c r} F(0,s) \\
\nonumber &- \frac{(s-c s+c z+c r z) }{cs} F(r,0) \\
\nonumber &-\frac{(c-1)^2 }{c^2} F(0,0)
}
\end{equation}
where the \emph{kernel}, $K(r,s)$, is
\begin{equation}
\label{3fw_kernel}
K(r,s) \equiv K(r,s;z) = 1-\frac{z(r+1) (s+1) (r+s)}{rs}.
\end{equation}	
\subsection{Symmetries and roots of the kernel}
\label{3fw_symmetries_and_roots_kernel}
We observe that the kernel $K(r,s)$ in \eref{3fw_main_fn_eqn_simplified_v1} and
\eref{3fw_kernel} is symmetric under the two transformations
\begin{equation}
\Phi: (r,s) \mapsto \left (s, r \right ), \qquad \Psi: (r,s) \mapsto \left ( r, \frac{r}{s} \right),
\end{equation}	 
where both $\Phi$ and $\Psi$ are involutions. These transformations generate a family of 12 symmetries $\mathcal{F}$ -- namely,

{\footnotesize
\begin{equation}
\fl
\mathcal{F} = \left \{ (r,s), (s,r), \left (r, \frac{r}{s} \right), \left (s, \frac{s}{r} \right ), \left (\frac{r}{s},r \right ), \left (\frac{s}{r},s \right), \left (\frac{r}{s}, \frac{1}{s} \right ), \left (\frac{s}{r}, \frac{1}{r} \right), \left (\frac{1}{s}, \frac{r}{s} \right), \left (\frac{1}{r}, \frac{s}{r} \right ), \left (\frac{1}{r}, \frac{1}{s} \right), \left (\frac{1}{s}, \frac{1}{r} \right)  \right \}.
\end{equation}
} 
\par
Now, considering $K(r,s)$ as a polynomial in $r$, we find two roots $\hat{s}_{\pm}$
\begin{equation}
\label{3fw_s_roots}
\eqalign {
\hat{s}_{\pm}(r;z) &= \frac{r - z \left(r^2 +2r+ 1\right)  \pm \sqrt{r^2-2 zr (1+r)^2 + z^2\left(r^2 - 1\right)^2 }}{2z(r+1)} , \\
\hat{s}_{-}(r;z)  &= (r+1)z+\frac{(r+1)^3 z^2}{r} + O(z^3), \\
\hat{s}_{+}(r;z) &= \frac{r}{(r+1) z}-(r+1)+ O(z).
}
\end{equation}
Moreover, as the kernel is symmetric in $r$ and $s$, we additionally have the roots $\hat{r}_{\pm}$
\begin{equation}
\label{3fw_r_roots}
\eqalign{
\hat{r}_{\pm}(s;z) &= \frac{s - z \left(s^2 +2s+ 1\right)  \pm \sqrt{s^2-2 zs (1+s)^2 + z^2\left(s^2 - 1\right)^2 }}{2z(s+1)},
}
\end{equation}
Providing $F \left (r,\hat{s}_{\pm} \right )$ remains in the ring of formal 
power series $\mathbb{Z}[r,a,c][[z]]$, substituting $s \mapsto \hat{s}_\pm$ into 
\eref{3fw_main_fn_eqn_simplified_v1} sets the kernel to zero and eliminates 
the left-hand side of the functional equation. Similarly, providing $F \left 
(\hat{r}_{\pm}, s \right )$ lies in $\mathbb{Z}[s,a,c][[z]]$, substituting $r 
\mapsto \hat{r}_\pm$ will set the left-hand side of the functional equation to 
zero.

By considering partial sums of $F(r,s)$ up to $O(z^n)$, one can show that 
$F \left (r,\hat{s}_- \right )$ and $F \left (\hat{r}_-,s \right )$ converge 
within the desired rings, while the other substitutions do not (essentially 
because $\hat{r}_+, \hat{s}_+ = O(z^{-1})$). Hence for the remainder of this 
paper, we define $\hat{s}(r;z) \equiv \hat{s}_{-}(r;z)$ and $\hat{r}(s;z) \equiv 
\hat{r}_{-}(s;z)$, and only consider substitution of these roots. 

Finally, when the kernel \eref{3fw_kernel} $K(r,s) = 0$ we have
\begin{equation}
rs = z (r+1)(s+1) (r+s) 
\end{equation}
and thus Lagrange inversion \cite{flajolet2009analytic} yields
\begin{equation}
\label{3fw_s_hat_r_hat_series_rep}
\eqalign {
{\hat{r}(s;z)}^k &= \sum_{n \geq k} \frac{k}{n} z^n (1+s)^n \sum_{j=k}^n {n \choose j}{n \choose j-k} s^{j-n}  \\
{\hat{s}(r;z)}^k &= \sum_{n \geq k} \frac{k}{n} z^n (1+r)^n \sum_{j=k}^n {n
\choose j}{n \choose j-k} r^{j-n}.
}
\end{equation}
Note that the above are closely related to the generating function of the 
Narayana numbers $N(n,j)=\frac{1}{n}{n\choose j}{n \choose j-1}$. This 
observation is sufficient to ensure that the above series contain only 
non-negative integer coefficients.

These explicit series representations for positive integer powers of the roots 
$\hat{r}$ and $\hat{s}$, will be used below to help find an explicit 
expression for the generating function.
\subsection{Using the symmetries of the kernel}
\label{3fw_obs_kernel_method}
%
Equipped with the roots $\hat{s}$ and  $\hat{r}$ as well as the family of symmetries $\mathcal{F}$ that leave the kernel invariant, we can now apply the obstinate kernel method. Specifically, we substitute $(r,s) \mapsto (r, \hat{s})$ and $(r,s) \mapsto (\hat{r}, s)$ into the simplified functional equation \eref{3fw_main_fn_eqn_simplified_v1}, subsequently applying a subset of transformations from $\mathcal{F}$ to generate a system of new functional equations. Mapping $(r,s) \mapsto (\hat{r}, s)$ we have the system
\par
{\scriptsize
\begin{equation}
\label{3fw_s_hat_1}
\fl
\eqalign {
0 = \frac{1}{c^2} - \frac{(\hat{r}-c \hat{r}+c z+ cs z)}{c \hat{r}} F(0,s) - \frac{(\hat{s}-c s+c z+c \hat{r} z) }{cs} F(\hat{r},0) -\frac{(c-1)^2 }{c^2} F(0,0), &\qquad (r,s) \mapsto (\hat{r}, s)
}
\end{equation}
\begin{equation}
\label{3fw_s_hat_2}
\fl
\eqalign {
0 = \frac{1}{c^2} - \frac{\left(\hat{r}-c \hat{r}+c z+\frac{c \hat{r} z}{s}\right)}{c r}  F \left(0,\frac{\hat{r}}{s} \right)  -\frac{s \left(\frac{\hat{r}}{s}-\frac{c \hat{r}}{s}+c z+c \hat{r} z\right)}{c \hat{r}} F(\hat{r},0) -\frac{(c-1)^2}{c^2} F(0,0), &\quad (r,s) \mapsto \left ( \hat{r}, \frac{\hat{r}}{s} \right)
}
\end{equation}
\begin{equation}
\label{3fw_s_hat_3}
\fl
\eqalign {
0 = \frac{1}{c^2}-\frac{s \left(\frac{1}{s}-\frac{c}{s}+c z+\frac{c \hat{r} z}{s}\right) }{c} F\left(0,\frac{\hat{r}}{s}\right) -\frac{s \left(\frac{\hat{r}}{s}-\frac{c \hat{r}}{s}+c z+\frac{c z}{s}\right)}{c \hat{r}} F\left(\frac{1}{s},0\right) -\frac{(c-1)^2}{c} F(0,0), &\quad (r,s) \mapsto \left ( \frac{1}{s}, \frac{\hat{r}}{s} \right)
}
\end{equation}
}
where the chosen subset of transformations guarantee that each functional 
equation \eref{3fw_s_hat_1} --- \eref{3fw_s_hat_3} \emph{only} contain 
non-negative powers of $\hat{r}$ and thus the generating functions are formally 
convergent in $\mathbb{Z}[c,s, \bar{s}][[z]]$. Considering the system of 
equations \eref{3fw_s_hat_1} - \eref{3fw_s_hat_3}, we can eliminate 
$F(\hat{r},0)$ by

{\footnotesize
\begin{equation}
\label{3fw_r_hat_simple_1}
\eqalign {
\nonumber 0 &= \left [ \mbox{coeff. of }F(\hat{r},0) \mbox{ in} \textrm{ 
\eref{3fw_s_hat_2}} \right ] \times \left [\textrm{RHS of \eref{3fw_s_hat_1}} 
\right ]\\ 
&
-  \left [ 
\mbox{coeff. of } F(\hat{r},0) \mbox{ in } \textrm{\eref{3fw_s_hat_1}} \right ] 
\times \left [\textrm{RHS of \eref{3fw_s_hat_2}}\right ] \\
&= - \frac{s \left(\frac{\hat{r}}{s}-\frac{c \hat{r}}{s}+c z+c \hat{r} z\right) }{c \hat{r}} \left [ \textrm{RHS of \eref{3fw_s_hat_1}} \right ]+ \frac{(s-c s+c z+c \hat{r} z) }{c s} \left [\textrm{RHS of \eref{3fw_s_hat_2}} \right ].
}
\end{equation}
}
In a similar vein we can eliminate $F \left (0, \frac{\hat{r}}{s} \right)$ from the system by

{\footnotesize
\begin{equation}
\label{3fw_r_hat_simple_2}
\eqalign {
\nonumber 0 &=\left [ \mbox{coeff. of }F \left (0, \frac{\hat{r}}{s} \right) \mbox{ in } \textrm{\eref{3fw_r_hat_simple_1}} \right ] \times \left [\textrm{RHS of \eref{3fw_s_hat_1}} \right ] \\
\nonumber &-  \left [ \mbox{coeff. of } F \left (0, \frac{\hat{r}}{s} \right) \mbox{ in } \textrm{\eref{3fw_s_hat_1}} \right ] \times \left [ \textrm{RHS of \eref{3fw_r_hat_simple_1}} \right] \\
\nonumber &= \frac{(s-c s+c z+c \hat{r} z) (-\hat{r} s+c \hat{r} s-c \hat{r} z-c s z) }{c \hat{r} s^2}\left [\textrm{RHS of \eref{3fw_s_hat_1}} \right ] \\
&- \frac{s \left(-\frac{1}{s}+\frac{c}{s}-c z-\frac{c \hat{r} z}{s}\right) }{c} \left [\textrm{RHS of \eref{3fw_r_hat_simple_1}} \right ]
}
\end{equation}
}
yielding a functional equation solely in terms of the generating functions $F(0,s), F \left (1/s ,0 \right )$ and $F(0,0)$. Specifically, we have
\begin{equation}
\label{3fw_r_hat_eqn_final_simple}
\fl
N_{1} (s, c;z) F \left (1/s,0 \right) + N_{2} (s, c;z) F(0,s) + N_{3} (s, c;z)  \left [ (c-1)^2 F(0,0) - 1\right ]  = 0
\end{equation}
where
\begin{equation}
\label{3fw_r_hat_eqn_final_simple_defn}
\fl
\eqalign {
N_1 (s, c;z) &= \frac{(s-c s+c (1+\hat{r}) z) (\hat{r}-c \hat{r}+c (1+s) z) (-(-1+c) \hat{r} s+c (\hat{r}+s) z)}{c^3 \hat{r}^2 s^2}, \\
N_2 (s, c;z) &= \frac{(\hat{r}-c \hat{r}+c s z+c \hat{r} s z) ((-1+c) \hat{r}-c (1+s) z) (1+c (-1+(\hat{r}+s) z))}{c^3 \hat{r}^2}, \\
N_{3} (s,c;z) &= \frac{(c-1)^2}{c^4} -\frac{(1+\hat{r}) (\hat{r}+s) \left(-1+\hat{r} s-s^3\right) z^2}{c^2 \hat{r} s^2} \\
\nonumber &-\frac{\left(-c \hat{r}+c^2 \hat{r}-c s+c^2 s-cs^2+c^2 s^2-c \hat{r} s^2+c^2 \hat{r} s^2\right) z}{c^4 \hat{r} s}.
}
\end{equation}
%
%

By an identical process, we can alternatively substitute in the root $\hat{s}$ along with a subset of transformations in $\mathcal{F}$
\begin{equation}
\left (r,s \right) \mapsto\left (r,\hat{s} \right), \left ( \frac{\hat{s}}{r}, \hat{s} \right), \left ( \frac{\hat{s}}{r}, \frac{1}{r} \right)
\end{equation}
that contain positive powers of $\hat{s}$ to yield an alternate refined functional equation containing $F(r,0), F(0, 1/r)$ and $F(0,0)$. However, under the horizontal reflection
\begin{equation}
(r,s, c) \mapsto (s, r, c)
\end{equation}
the generating $F(r,s)$ along with any corresponding functional equations will be invariant and it thus suffices to solely consider \eref{3fw_r_hat_eqn_final_simple} for our subsequent analysis.
\par
By using the symmetries of the kernel we have established the refined functional equation \eref{3fw_r_hat_eqn_final_simple} containing unknown generating functions in only the catalytic variable $s$. The potential benefit of this new equation is that by extracting the coefficients of $s^i$ for some choice of $i$ we hope to establish a relation solely in terms of $F(0,0,c;z) \equiv G(c,1;z)$. This is indeed what we will proceed to do in \Sref{3fw_soln_eql_case}.
\section{Solving the generating function}
\label{3fw_soln_eql_case}
Our aim is to utilise the refined functional equations that was established in 
\Sref{3fw_obs_kernel_method}. We start by dividing 
equation \eref{3fw_r_hat_eqn_final_simple} by the coefficient 
of $F(0,s)$, namely $N_2(s,c;z)$. It was then observed that $N_1/N_2$ was 
actually a rational function of $s,c$ and $z$. Multiplying through by the 
associated denominator then gives:
{\normalsize
\begin{equation}
\label{3fw_r_hat_eqn_final_simple_eql_case}
\fl
\eqalign {
\nonumber &s-c s+c z+(-1+c) s^2 (-1+c+c z) F \left (1/s,0 \right) \\
\nonumber &- s \left(1+s+c \left(-2+c-s+\left(-1+c+s^2\right) z\right)\right) F(0,s) \\
&= -s \left(1+s+c \left(-2+c-s+\left(-1+c+s^2\right) z\right)\right) \frac{N_{3} (s,c;z)}{N_{2} (s,c;z)}  \left [1- (c-1)^2 F(0,0) \right ],
}
\end{equation}
}
where our algebraic functions $N_i(s,c;z)$ are defined in \eref{3fw_r_hat_eqn_final_simple_defn}. Extracting the coefficient of $s^1$ of \eref{3fw_r_hat_eqn_final_simple_eql_case} gives us

{\normalsize
\begin{equation}
\label{3fw_r_hat_eqn_final_simple_eql_case_2}
\fl
\eqalign {
\nonumber &1-c-(c-1) (cz+c-1) F(0,0)+ (c-1) (cz+c-1) [s^1] F(s,0) \\
&= -[s^1] \left \{ s \left(1+s+c \left(-2+c-s+\left(-1+c+s^2\right) z\right)\right) \frac{N_{3} (s,c;z)}{N_{2} (s,c;z)} \right \}  \left [1- (c-1)^2 F(0,0) \right ].
}
\end{equation}
}
Finally, to eliminate the boundary term $[s^1] F(s,0)$, we recall relation \eref{3fw_F00_fn_eqn} and observe that our walks are symmetric under a horizontal reflection. Thus, $[r^1] F(r,0) = [s^1] F(0,s)$ (more generally $F(r,s) = F(s,r)$) and so we have the relation
\begin{equation}
\label{3fw_eql_case_f00}
[s^1] F(s,0) = -\frac{1}{2 c^2 z}-\frac{\left(-1+2 c^2 z\right) F(0,0)}{2 c^2 z},	
\end{equation}	
which when subsequently substituting into \eref{3fw_r_hat_eqn_final_simple_eql_case_2} yields an expression for $F(0,0)$
{\small
\begin{equation}
\fl
\label{3fw_G1c_expr}
G(c,1) = F(0,0) = \frac{1}{(c-1)^2} \left (1 + \frac{c (2 - 3 c + c^2 - 4 c z + c^2 z + 2 c^3 z + 4 c^2 z^2)}{1-c+c \left(-1-4 c+4 c^2\right) z-2 (c-1) c^2 z [s^1]H(c;z) + 4 c^3 z^2}\right )
\end{equation}
}
where
\begin{equation}
\label{3fw_eql_case_H}
H(c;z) \equiv - s \left(1+s+c \left(-2+c-s+\left(-1+c+s^2\right) z\right)\right) \frac{N_{3} (s,c;z)}{N_{2} (s,c;z)}.	
\end{equation}
What remains is to explicitly extract the coefficient of $s^1$ from $H(r,c;z)$. We begin by expanding $H(c;z)$ as a power series in $c$ giving us
\begin{equation}
\label{3fw_eql_case_H_series_in_c}
H(c;z) = \left [ s^4 z +2s^3 z + s^2 \left(2z-1\right)+sz +z \right ] \sum_{m \geq 0} c^m X(c;z)^m
\end{equation}
where
\begin{equation}
\label{3fw_Xm}
\eqalign {
\nonumber X(c;z)^m &= \frac{z^m (s+1)^{m-1}}{s^{m+1}} \left [\hat{r} + (1+s) \right]^m \\
&= \frac{z^m}{s^{m+1}} \sum_{k=0}^m {m \choose k} \hat{r}^k (s+1)^{2m-k-1}
}
\end{equation}
Using our expansion for $\hat{r}^k$ in \eref{3fw_s_hat_r_hat_series_rep} we find
\begin{equation}
\label{3fw_Xm_v2}
\eqalign {
X(c;z)^m &= \sum_{k=0}^m {m \choose k} \sum_{n \geq k} \frac{k}{n} z^{m+n} (s+1)^{2m+n-k-1} \sum_{j=k}^n {n \choose j}{n \choose j-k} s^{j-m-n-1}.
}
\end{equation}
Now, from \eref{3fw_eql_case_H_series_in_c}, $[c^m s^1]H(c;z)$ is
\begin{equation}
\label{3fw_eql_case_HcMs1}
[c^m s^1] H(c;z) = \left (z[s^{-3}] + 2z[s^{-2}] + \left(2z-1\right)[s^{-1}] + z [s^0]  + z [s^1] \right ) X(c;z)^m.
\end{equation}
Thus, equipped with our expansion for $X(c;z)^m$ \eref{3fw_Xm_v2}, one can explicitly find $[c^m s^1] H(c;z)$ by extracting and subsequently combining the coefficients $[s^i] X(c;z)^m$ for $i=-3,-2, \dots,1$. For instance, we find $[s^0] X(c;z)^m$
\begin{equation}
\label{3fw_s0Xm}
\eqalign {
\nonumber [s^0] X(c;z)^m &= \sum_{k=1}^m {m \choose k} \sum_{n \geq k} \frac{k}{n} z^{m+n} \sum_{j=k}^n {2m+n-k-1 \choose m+n-j+1} {n \choose j} {n \choose j-k} \\
&+z^{m} {2m-1 \choose m+1}.
}
\end{equation} 
Therefore, along with $[s^0] X(c;z)^m$ found in \eref{3fw_s0Xm} there remain four other components in \eref{3fw_eql_case_HcMs1} whose series representation can be determined in the same fashion, giving us an expansion for $[s^1] H(c;z)$ as a series in $c$. Finally, we change the order of summation to get terms that are a power series in $z$ and with the aid of \verb+Maple+ \cite{maple_package} to combine our sums we find the exact solution for $[s^1] H(c;z)$ to be
\begin{equation}
\label{3fw_s1H}
\eqalign {
&[s^1] H(c;z) = z + \frac{1-b}{b} + \frac{1-2cz-c^2z + \left(c^2z - 2c^2 z^2 -1 \right) \sqrt{1-4cz}}{2c^2z \sqrt{1-4cz}} \\
&+ J(c;z)
}
\end{equation}
where $J(c;z)$ is

{\footnotesize
\begin{equation}
\label{3fw_J}
\fl
\eqalign {
\nonumber J(c;z) = \sum_{i \geq 3} z^i \sum_{m=1}^{i-1} c^m \sum_{k=1}^{i-m-1} {m \choose k} \sum_{j=k}^{i-m-1} &\left \{\frac{k}{i-m-1} {i-m-1 \choose j} {i-m-1 \choose j-k} \left [ {m+i-k \choose i-j} + {m+i-k \choose i-j-2} \right] \right . \\
\nonumber & - \left . \frac{k}{i-m} {i-m \choose j} {i-m \choose j-k} {m+i-k-1 \choose i-j-1} \right \} \\
& - \sum_{i \geq 2} z^i \sum_{m=1}^{i-1} c^m \sum_{k=1}^{i-m} {m \choose k} \frac{k}{i-m} {i-m \choose i-k-m} {m+i-k-1 \choose m-1}
}
\end{equation}
}
Substituting \eref{3fw_s1H} into \eref{3fw_G1c_expr} yields
\begin{equation}
\label{3fw_full_soln_gc1}
G(c,1;z) = \frac{1}{c(c-2) + 1} \left (1+ \frac{\left [  c \left(2-3 c+c^2\right)+ z c^3 \right ] \sqrt{1-4cz}}{G_b(c,1;z)} \right )
\end{equation}
where
{\small
\begin{equation}
\label{3fw_Gc1denom}
\fl
G_b(c,1;z) = (1-c) \left(-1+2 c z+c^2 z\right) +c z \sqrt{1-4 c z} \left [-1+c-c^2+2c(c-1) J(c;z) \right ]. 
\end{equation}
}
Finally recalling relation \eref{3fw_full_model_decomp} between  $G(c,d;z)$ and $G(c,1;z)$, our solution of the full model is
\begin{equation}
\label{3fw_full_model_expr}
G(c,d;z) = \frac{1}{cd(c-2)+ 1} \left (1+ \frac{\left[c d (c-2) (c-1)^2 +c^3 d z (c-1) \right] \sqrt{1-4 c z}}{G_b(c,d;z)} \right ),
\end{equation}
where

{\small
\begin{equation}
\label{3fw_Gc1denom_full}
\fl
\eqalign {
\nonumber G_b(c,d;z) &= 1-2 cd+c^2d-c (c+2) \left(1-2 cd+c^2 d\right) z \\
&+\sqrt{1-4cz} \left[c (2-c)  (d-1)+cz \left(1-2cd+2c^2d-c^3d\right)+ 2 zc^2 \left(1-2 c d+c^2 d\right)  J(c;z)\right].
}
\end{equation}
}
\section{Phase structure and transitions}
\label{3fw_analysis}
\subsection{Singularity structure of $G(c,1)$}
\label{3fw_sing_analysis_c1}
Recall from \Sref{3fw_soln_eql_case} that our exact solution to $G(c,1)$ was expressed as
\begin{equation}
G(c,1;z) = \frac{1}{c(c-2) + 1} \left (1+ \frac{\left [  c \left(2-3 c+c^2\right)+ z c^3 \right ] \sqrt{1-4cz}}{G_b(c,1;z)} \right )
\end{equation}
with the denominator $G_b(c,1;z)$ defined in \eref{3fw_Gc1denom}. In particular $G_b(c,1;z)$ is an expression in terms of $z,c$ and the power series $J(c;z)$ which itself is defined at \eref{3fw_J}. Hence the dominant singularity $z_s(c,1)$ of our generating function is dependent on the dominant singularity $z_J(c,1)$ of $J(c;z)$ along with any poles that arise from the roots of $G_b(c,1;z)$.
Now, by an exact approach featured in \cite{tabbara2012pulling}, we employ the 
technique of differentiating hyperexponential functions under the integral sign 
to determine a linear homogenous differential equation with polynomial 
coefficients in $z$ and $c$ that is satisfied by the series. Once again, we 
utilise the \verb+Maple+ package \verb+DETools+ which implements the so-called 
`fast' Zeilberger algorithm applicable to hyperexponential functions 
\cite{almkvist1990method}, to find the linear differential operator 
$\mathcal{L}$ where 

{ \small
\begin{equation}
\label{3fw_final_short_DE_J}
\fl
\eqalign {
\nonumber \mathcal{L}&= \left [ 16384 c^{15} \left ( 279 c^{6} \dots + 23808 
\right ) z^{26} + \dots + \left(-10 c^{17}+ \dots + 4 \right) z^3 \right ] \left 
( {\partial/\partial_{z}}\right )^7 \\
\nonumber &+\left [23592960 c^{15} \left ( 279 c^{6} \dots + 23808 \right ) 
z^{20} + \dots + \left(-302400 c^{18}+ \dots -72000 c^2 \right) \right ] 
\left({\partial/\partial_{z}}\right), \\
}
\end{equation}
}
satisfying the equation
\begin{equation}
\label{3fw_J_DE_homogeneous}
\mathcal{L} J(c;z) = 0.
\end{equation}
In \ref{appendix:3fw_g1c_DE_leading_coeff} we explicitly write out the leading 
polynomial coefficient of \eref{3fw_final_short_DE_J} whose zeroes correspond to 
the singularities of $J(c;z)$ which allows us to determine the dominant 
singularity structure $z_J(c,1)$ of $J(c;z)$
\begin{equation}
\label{3fw_sing_J}
z_J(c,1) =
\cases{
z_b \equiv 1/8, & $c \leq 4/3$ \\
z_{c}(c) \equiv \frac{1-c+\sqrt{c^2-c}}{2c},& $c > 4/3$.
}
\end{equation}
Thus we find a critical point at $c=4/3$ when both the singularities $z_b = 
z_{c}(4/3) = 1/8$ coincide. Note that the above implies that the square-root 
singularity of $G(c,1;z)$ at $z=1/4c$ is subdominant. Recall further that with 
our differential equation one can also determine the corresponding linear 
recurrence for the coefficients $J_n$ of $J(c;z)$ and in particular with the 
assistance of the \verb+Maple+ package \verb+Gfun+ \cite{SaZi94} we find
\par
\begin{figure}[h]
\psfrag{Z}{\small{$z$}}
\psfrag{T}{\small{$z$}}
\psfrag{C}{\small{$c$}}
\centering
\includegraphics[width=200px]{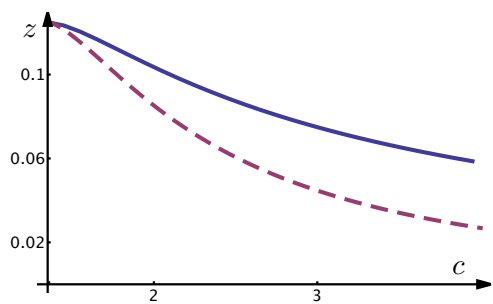}
\caption[The singularities $z_p(c)$ and $z_c(c)$ for $c > 4/3$]{The singularities $z_p(c)$ (dashed) and $z_c(c)$ (solid) for $c > 4/3$}
\label{fig:3fw:zp_vs_zc_sing} 
\end{figure}

{ \small
\begin{equation}
\label{Jrecurrence}
\fl
\eqalign {
\nonumber &-6120576000 (c-1)^{15} \left(2-10 c+5 c^2\right) J_{n+23} \\
\nonumber &+ q_2(c,n) J_{n+22}+ q_3(c,n) J_{n+21} + q_4(c,n) J_{n+20} + \dots \\
&+ 1966080 c^{15} \left(23808-73632 c+82960 c^2-35756 c^3-2310 c^4+4358 c^5+279 
c^6\right) J_n = 0,
}
\end{equation}
}
where $q_{i}(c,n) \in \mathbb{Z}[c,n]$, giving us a homogeneous linear 
recurrence equation of order 23 with polynomial coefficients in $n$.
The growth of the coefficients $J_n$ of $J(c;z)$ can be directly determined from 
recurrence \eref{Jrecurrence} by appealing to the method of Wimp and Zeilberger 
\cite{wimp1985resurrecting}, showing the existence and specific form of a basis 
set of asymptotic solutions for any given linear recurrence which, in 
particular, contains rational coefficients (in $n$). In this instance, we 
substitute into \eref{Jrecurrence} the ansatz
\begin{equation}
\label{tfw_ansatz_coeff_g1c}
J_n = b_0 \mu^n n^{\gamma -1}, \quad b_0 \ne 0
\end{equation}
where $\mu, \gamma, b_{0} \in \mathbb{R}$. By collecting dominant powers of $n$ 
and equating their corresponding coefficients to zero we can solve for  $\mu$, 
$\gamma$. In doing so we find
\begin{equation}
\label{3fw_Jcoeff_growth}
J_n \equiv J_n(c) = 
\cases {
B_{-} 8^n n^{-4}, & $c < 4/3$, \\
B_0 8^n n^{-2}, & $c=4/3$, \\
B_{+} z_c(c)^{-n} n^{-1/2}, & $c > 4/3$
}
\end{equation}
and moreover that the singular part of the generating function near the radius 
of convergence behaves as
\begin{equation}
\label{3fw_J_growth}
J(c;z) \sim
\cases{
B_{-}(1-8z)^3 \log (1- 8z), & $c < 4/3$, \\
B_0 (1-8z)\log(1- 8z), & $c=4/3$, \\
B_{+} (1 - z/z_c (c))^{-1/2},& $c > 4/3$.
}
\end{equation}
Finally, we want to determine whether $G(c,1;z)$ exhibits any additional 
critical points that arise from the smallest real and positive root $z_p(c)$ of 
$G_b(c;z)$. By taking a truncated series approximation of $J(c;z)$, we can 
numerically find the roots of $G_b(c;z)$ and estimate $z_p(c)$ which we plot in 
\Fref{fig:3fw:zp_vs_zc_sing}, suggesting that $z_p(c) < z_c(c)$ for all $c>4/3$. 
We know that as $c \rightarrow \infty$, $G(c,1;z)$ is dominated by those 
configurations where all three walks coincide for every step so that
\begin{equation}
G(c,1;z) \sim \frac{1}{1-2c^2z}, \quad c \rightarrow \infty.
\end{equation}
%
%
%
Hence, we indeed should expect the pole $z_p(c)$ to dominate $z_c(c)$ for large $c > 4/3$. Now, we need to justify that $z_p(c) < z_c(c)$, for \emph{all} $c > 4/3$. If we assume the converse, then there must exist $c^{\star} > 4/3$ such that $z_p(c^{\star}) = z_c(c^{\star})$. From \eref{3fw_J_growth}, the analytic and singular expansion of $J(c^{\star};z)$ is given as
\begin{equation}
J(c^{\star};z) \sim A_{+} + B_{+} (1 - z/z_c (c^{\star}))^{-1/2}, \quad z \rightarrow z_c (c^{\star}), A_{+},B_{+} \ne 0,
\end{equation}
and our expansion for $G_b(c^{\star}, 1; z)$ becomes
\begin{equation}
\fl
G_b(c^{\star}, 1; z) \sim f_0 (c^{\star}) + \frac{f_1(c^{\star}) B_{+}}{\sqrt{1 - z/z_c (c^{\star})}} + f_2(c^{\star}) B_{+} \sqrt{1 - z/z_c (c^{\star})}, \quad z \rightarrow z_c (c^{\star}),
\end{equation}
where $f_i(c^{\star})$ are algebraic functions in $c^{\star}$. Note that $A_{+}$ has been absorbed into $f_0$. Now, as $G_b(c^{\star}, 1; z_p(c^{\star})) \sim 0$, we require $f_0 = f_1 = 0$, however considering $f_1(c)$, where
\begin{equation}
f_1(c) = \frac{2 c^2 (c-1)^2  \left(2-3 c+3 \sqrt{c(c-1)}\right)}{\sqrt{-1+2c-2 \sqrt{c(c-1)}}},
\end{equation}
we find the only roots of $f_1$ are at $c=0,1$ and $4/3$. Thus, there can not exist $c^{\star} > 4/3$ such that $z_c(c^{\star}) = z_p(c^\star)$.
With that in mind, we can finally conclude that the dominant singularity, 
$z_s(c,1)$, of $G(c,1;z)$ is
%
%
\begin{equation}
\label{3fw_sing_G_c_1}
z_s(c,1) =
\cases {
z_b \equiv 1/8, & $c \leq 4/3$, \\
z_{p}(c), & $c > 4/3$
}.
\end{equation}
%
%
\subsection{Phase transitions of $G(c,1)$}
\label{trans:gc1}
Since the dominant singularity of $G(c,1)$ contains a single non-analytic point, we 
would expect that our model exhibits two distinct phases. To begin to 
characterise these phases we introduce the order parameter $\mathcal{C}(c)$ 
denoting the limiting average number of shared contact sites:
\begin{equation}
\mathcal{C}(c) = \lim_{L \rightarrow \infty} \frac{\langle m_c \rangle}{L} 
= c \frac{\partial}{\partial c} \log{z_s(c,1)}.
\end{equation}
The system is in a \emph{free} phase when
\begin{equation}
\mathcal{C} =0,
\end{equation}
while a \emph{gelated} phase is observed when
\begin{equation}
\mathcal{C} > 0.
\end{equation}
Recall from \sref{3fw_sing_analysis_c1} that $z_s(c,1) = z_b \equiv 1/8$ for $c \leq 4/3$, implying that $\mathcal{C} = 0$ over the same region.
For $c \geq 4/3, z_s(c,1) = z_p(c, 1)$, which is given implicitly as the smallest positive root of the expression

{\footnotesize
\begin{equation}
\label{ch5:3fw_Gc1denom_v2}
\fl
G_b(c,1;z) = -1-c^2 z-c^3 z+c (2 z + 1)+\sqrt{1-4cz} \left[-c z+c^2 z-c^3 z+\left(-2 c^2 z+2 c^3 z\right) J(c;z) \right]. 
\end{equation}
}

\begin{figure}[h]
\psfrag{O}{\small{$\mathcal{C}$}}
\psfrag{C}{\small{$c$}}
\centering
\includegraphics[width=200px]{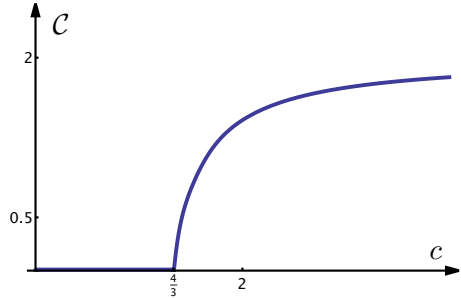}
\caption[The limiting average number of shared contacts $\mathcal{C}$ when 
$d=1$]{The limiting average number of shared contacts $\mathcal{C}$ when $d=1$. 
The system exhibits a second-order transition at $c=4/3$ respectively}
\label{fig:3fw:ordC_d1} 
\end{figure}

Now, consider the expansion of $G_b(c,1;z_p(c)) = 0$ around $c=4/3$. With the aid of the \verb+Maple+ \cite{maple_package},
we explicitly compute $J(4/3;1/8)$ where
\begin{equation}
J(4/3;1/8) = \frac{-24+13 \sqrt{3}}{8 \sqrt{3}} \approx -0.107051,
\end{equation}
and thus as our expansion of $J(c;z_p)$ is given as
\begin{equation}
J(c;z_p) \sim \frac{-24+13 \sqrt{3}}{8 \sqrt{3}} + B_0(1-8z_p) \log(1-8z_p), \quad c \downarrow 4/3,
\end{equation}
so that
\begin{equation}
\label{G_B_expansion_z_p}
\fl
G_b(c,1;z_p) \sim g_0(c-4/3) + \left[g_1 + g_2(c-4/3) \right] (1-8z_p) \log(1-8z_p), \quad c \downarrow 4/3, g_i \ne 0.
\end{equation}
As $G_b(c,1;z_p) \sim 0$, we apply the implicit function theorem to solve for the dominant behaviour of $\partial z_p / \partial c$ in \eref{G_B_expansion_z_p} to find
\begin{equation}
\eqalign {
\frac{\partial}{\partial c} z_p(c,1) &\sim \frac{3 \left(45-22 \sqrt{3}+72 B (1-8 z_p) \log(1-8 z_p)\right)}{64 B (-32+27 c) (1+\log(1-8 z_p))}, \\
& \sim 0, \quad c \downarrow 4/3
}
\end{equation}
Hence we deduce that our order parameter $\mathcal{C}(c)$ is continuous at $c=4/3$. In particular, we find the dominant behaviour second-order derivative of $z_p$ is given as
\begin{equation}
\frac{\partial^2}{\partial c^2} z_p(c,1) \sim \frac{81 \left(-45+22 \sqrt{3}\right)}{1024 B}, \quad c \downarrow 4/3
\end{equation}
and we conclude that we observe a second-order phase transition with a 
finite-jump discontinuity in the first-derivative of $\mathcal{C}(c)$ as seen in 
\Fref{fig:3fw:ordC_d1}. Note, that as $c \rightarrow \infty$, free energy is 
minimised for those configurations where all three walks are zipped together and 
the total weight of such a configuration of length $N$ will be $c^{2N}$. Thus 
$\mathcal{C}(c) \rightarrow 2$ as $c \rightarrow \infty$ which is indeed what we 
observe in \Fref{fig:3fw:ordC_d1}.

%

%
\subsection{Analysis and phase diagram of the full model}
\label{3fw_sing_analysis_c_full}
From our analysis in \sref{3fw_sing_analysis_c1}, the dominant singularity 
$z_s(c,d)$ of the generating function can be one of either $z_b \equiv 1/8$ or the pole $z_p(c,d)$ which is now a function of \emph{both} $c$ and 
$d$. Substituting $z_b$ and $c=4/3$ to locate the zero of $G_b(4/3,d;z_b)$, we 
find that the two singularities coincide when $d=9/8$ and further that $z_p(4/3, d)$ is a strictly decreasing function of $d$ for all $d>9/8$.
\par
\begin{figure}[h]
\psfrag{D}{\small{$d$}}
\psfrag{C}{\small{$c$}}
\psfrag{A}{\small{$\alpha(d)$}}
\psfrag{B}{\small{$\beta(c)$}}
\centering 
\includegraphics[width=300px]{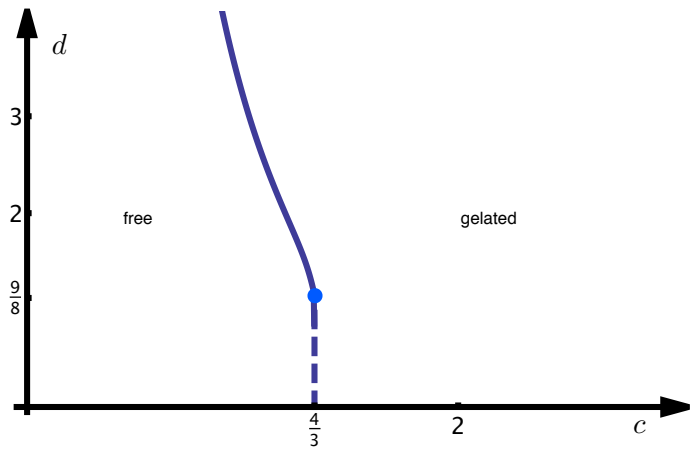} 
\caption[Three friendly interacting walks in the bulk: phase diagram]{The phase diagram of our full model. First and second-order transitions are indicated by solid and dashed lines respectively. All phase boundaries coincide at $c=4/3$ and $d=9/8$.} 
\label{3fw:fig:phase_diag} 
\end{figure}

What remains is to determine the dominant singularity over the region $c<4/3,d >0$. Overall, estimating the location of $z_p(c,d)$ over this region we find
\begin{equation}
\label{3fw_sing_full}
z_s(c,d) =
\cases {
z_b \equiv 1/8, & $c \leq 4/3, d < 9/8$ \\
z_b, & $c \leq \alpha(d), d \geq 9/8$ \\
z_{p}(c,d), & $c > 4/3, d < 9/8$ \\
z_{p}(c,d), & $c > \alpha(d), d \geq 9/8$ 
}
\end{equation}
where the boundary $\alpha(d)$ corresponds to when the singularities $z_p(c,d)=z_b$ coincide respectively.
With the full dominant singularity structure established, our system exhibits the same phase regions as per the $d=1$ model --- namely, a free phase is observed when
\begin{equation}
\mathcal{C} = 0,
\end{equation}
while our system is in a gelated phase when
\begin{equation}
\mathcal{C} > 0.
\end{equation}
Equipped with the phases of our system we plot the phase diagram in 
\Fref{3fw:fig:phase_diag}.
%
%
By a similar argument employed in \sref{trans:gc1}, for general $d<9/8$ we observe a \emph{second-order} transition when moving from a free to gelated phase. Now, considering the boundary $\alpha(d)$ where $d> 9/8$, we note that our our boundary lies in the region $c<4/3$, and hence the expansion of $J(c;z_p)$ is given as
\begin{equation}
\label{j_z_p_expansion}
J(c; z_p) \sim A_{-} + B_{-} (1-8z_p)^3\log(1-8z_p), \quad c \rightarrow \alpha(d).
\end{equation}
Substituting \eref{j_z_p_expansion} into $G_b(c,d;z_p) = 0$, we again apply the implicit function theorem to solve for the dominant behaviour of $\partial z_p(c;d) / \partial c$ as $c \downarrow \alpha(d)$ to find
\begin{equation}
\fl
\eqalign {
&\frac{\partial}{\partial c} z_p(c,d) \sim -\frac{17}{8 \alpha } -\frac{15 A_{-}}{4}+\frac{3}{8} (-7+10 d) \\
\nonumber &+\left(-\frac{15 {A_{-}}^2}{2}+\frac{1}{16} A_{-} (-203+352 d)+\frac{1}{16} \left(-77+122 d-120 d^2\right)\right) \alpha + O(\alpha^2), \quad c \rightarrow \alpha(d).
}
\end{equation}
Relying on our estimates for $A_{-}$ along $\alpha(d)$, we find that $\partial z_p(c;d) / \partial c$ is non-zero and hence deduce that our order parameter $\mathcal{C} > 0$ as $c \downarrow \alpha(d)$. Thus, for $d > 9/8$, the system exhibits a \emph{first-order} transition when moving from a free to gelated phase.
Moreover, employing a low-temperature argument, we can determine asymptotics for the boundary
$\alpha(d)$. Specifically, as $d \rightarrow \infty$, $G(c,d;z)$ is dominated by those configurations where all three walks 
coincide for every step --- that is,
\begin{equation}
G(c,d;z) \sim \frac{1}{1-2 c^2 d z}, d \rightarrow \infty,
\end{equation}
and equating the singularity of the limiting generating function with $z_b$, we find
\begin{equation}
\alpha(d) \sim 0, \quad d \rightarrow \infty.
\end{equation}
Finally, to describe the singular behaviour of $G(c,d)$ across all phases and boundaries, we consider the expansion of the generating function around the dominant singularity $z_s(c,d)$. In particular, the singular behaviour of $G(c,d)$ is driven by the expression $G_b(c,d)$, and hence $J(c,d)$. Recall in \sref{3fw_sing_analysis_c1} that the singular behaviour of $J(c,d)$ was determined in \eref{3fw_Jcoeff_growth}. With that in mind, we find that for the free phase
\begin{equation}
\eqalign {
G_b(c,d;z) \sim f_0(c,d) + f_1(c,d) (1-8z)^3 \log (1- 8z), &\quad f_i \ne 0,
}
\end{equation}
and thus,
\begin{equation}
\eqalign {
G(c,d;z)_{singular} \sim D_{-} (1-8z)^3 \log (1- 8z), &\quad z \rightarrow \frac{1}{8}, D_{-} \ne 0.
}
\end{equation}
In a similar fashion, in the gelated phase where $z_p(c,d)$ is dominant, $J(c,d;z_p)$ is analytic and hence
\begin{equation}
G(c,d;z)_{singular} \sim D_{+} \left (1 - z/z_p(c,d) \right)^{-1}, \quad z \rightarrow z_p(c,d), D_{+} \ne 0.
\end{equation}
Now, considering the singular behaviour of $G(c,d)$ along the phase boundaries requires a bit more care. Fixing $c=4/3$, we recall from our analysis in \sref{trans:gc1} that
\begin{equation}
J(4/3;1/8) = \frac{-24+13 \sqrt{3}}{8 \sqrt{3}},
\end{equation}
implying that
\begin{equation}
\label{3fw:J_expansion_4/3}
J(4/3;z) \sim \frac{-24+13 \sqrt{3}}{8 \sqrt{3}} + B_0 (1-8z)\log(1-8z), \quad z \rightarrow \frac{1}{8}.
\end{equation}
Along the free to gelated boundary where $c=4/3, d < 9/8$, we substitute our expansion for $J(4/3;z)$ in \eref{3fw:J_expansion_4/3} into the generating function and find
\begin{equation}
\fl
\eqalign {
G(4/3,d;z)_{singular} &\sim  \frac{-18d \left(1- 8z\right)}{(9 - 8 d)^2 \left(24-13 \sqrt{3}+8 \sqrt{3} \left [ \frac{-24+13 \sqrt{3}}{8 \sqrt{3}}+B_0 (1-8z) \log{1-8 z} \right ] \right)} \\
&\sim -\frac{18 d}{B_0 (9 - 8 d)^2 \log{(1 - 8 z)}}, \quad z \rightarrow \frac{1}{8}.
}
\end{equation}

At the point $c=4/3$, $d=9/8$, where the two singularities $z_b$ and $z_p(4/3, 9/8)$ coincide, we have the distinct expression for $G(4/3,9/8;z)$ that arises from a simplification in the generating function $G(4/3,1;z)$ and hence our primitive generating function $P(4/3;z)$, where
\begin{equation}
G(4/3,9/8;z) = \frac{32 z J(4/3;z) - 8+\frac{9}{\sqrt{1-\frac{16 z}{3}}}-4z \left(-3+\frac{10}{\sqrt{1-\frac{16 z}{3}}}\right)}{1-8 z},
\end{equation}
and further we find that the singular behaviour of our generating function at the critical point is
\begin{equation}
G(4/3,9/8;z)_{singular} \sim  D_{\star} \log \left ( 1-8z \right), \quad z \rightarrow \frac{1}{8}, D_{\star} \ne 0.
\end{equation}
Finally, considering the boundary $\alpha(d)$, we return to our original expression for the full generating function in \eref{3fw_full_model_expr} where $G_b \sim 0$ and along $\alpha(d)$ we find
\begin{equation}
\eqalign {
G(\alpha(d),d;z)_{singular} &\sim \frac{D_{\alpha}}{(1-8z) \left [f_0 + f_1 (1-8z)^2 \log(1-8z) \right]}, \quad D_{\alpha},f_i \ne 0 \\
&\sim \frac{D_{\alpha}}{(1-8z)}, \quad z \rightarrow \frac{1}{8}.
}
\end{equation}
Once equipped with the singular behaviour of $G(c,d;z)$, we can readily obtain the growth rate of the coefficients $Z_n \equiv Z_n(c,d)$ along the entire phase space, which we summarize in \tref{3fw_coeff_growth_table}.
\begin{table} \caption{\label{3fw_coeff_growth_table} The growth rates of the coefficients $Z_n(c,d)$ modulo the amplitudes of the full generating function $G(c,d;z)$ over the entire phase space. }
\begin{center}
\begin{tabular}{l | c} phase region &$Z_n(c,d) \sim $\\
\hline
free & $8^n n^{-4}$\\ 
gelated & $z_p(c,d)^{-n} n^{0}$ \\
\hline
free to gelated boundary, $d < 9/8$ & $8^n n^{-1} \log n$\\
free to gelated boundary, $d > 9/8$ & $8^n n^{0}$\\
\hline
$c=4/3, d=9/8$ & $8^n n^{-1}$\\
\hline 
\end{tabular} 
\end{center}
\end{table}
\subsection{A change of parameters: $t \equiv dc^2$}
\label{3fw_analysis_isolated_parameters}
\begin{figure}[h]
\psfrag{D}{\small{$t$}}
\psfrag{C}{\small{$c$}}
\centering 
\includegraphics[width=250px]{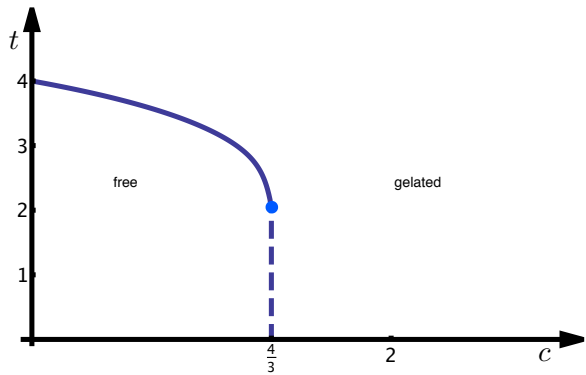} 
\caption[The phase diagram of the model when setting $d=t/c^2$]{The phase diagram of our full model when setting $d=t/c^2$. First and second-order transitions are indicated by solid and dashed lines respectively. All phase boundaries coincide at $c=4/3$ and $t=2$.} 
\label{3fw:fig:phase_diag_t} 
\end{figure}

Recall from our original model that for a given configuration when all three 
walks coincide on a site we incorporate both double and triple shared effects, 
weighting that site by $c^2d$. As a reparametrization of this model, we can 
define the parameter $ t \equiv c^2d$, thereby isolating the effects of double 
and triple shared contacts which now have corresponding weights $c$ and $t$ 
respectively. This allows us to consider a system of three interacting polymers 
where the energy required to graft all three or just two polymers are 
independent. From \eref{3fw_full_model_decomp} we can immediately express the 
generating function $G(c,t;z)$ as
\begin{equation}
\label{3fw_full_model_decomp_v2}
G(c,t;z) = \frac{G(c,1;z)}{\frac{t}{c^2} \left [1-G(c,1;z) \right ] + G(c,1;z)}
\end{equation}
where the exact solution to $G(c,1;z)$ has been previously established in 
\eref{3fw_full_soln_gc1}. Thus the singularity structure of our model 
\eref{3fw_sing_full} remains unchanged except for a rescaling of the pole $z_p$ 
and all singularities coincide at $c=4/3, t=2$. More generally the full phase 
diagram is presented in \Fref{3fw:fig:phase_diag_t}.
%
%
At $c=0$, the series $J(0;z) = 1$ and we find the denominator $G_b(0,t;1/8) = 1 
- t/4$. Thus we have a critical point at $t=4$, which is precisely what we 
observe in \Fref{3fw:fig:phase_diag_t}.
\section{Conclusion}
We have solved a model of three interacting friendly directed walks in the 
bulk. The system has two distinct interaction parameters $c$ and $d$,
corresponding to double and triple shared-contact sites to capture the effects 
of gelation. We established a 
combinatorial decomposition for the model's full generating function in terms of 
the corresponding simplified generating function (when $d=1$). We then 
derived a functional equation for this simpler generating function, and by 
means of the obstinate kernel method, proceeded to solve for both $G(c,1;z)$, 
and subsequently, the full generating function $G(c,d;z)$.

Our analysis of the simplified generating function where $d=1$ 
showed the existence of two phases which we classified as 
\emph{free} and \emph{gelated}, exhibiting a second-order phase transition. 
We then analysed the full model, presenting the phase diagram and showing that the phase space remains partitioned into two distinct phases. In particular, we located second and first-order phase boundaries, which coincide at $c=4/3, d=9/8$. 

It may be natural to consider next a more general asymmetric model incorporating separate 
parameters for top to middle and middle to bottom shared contacts. We have 
attempted to analyse this more general model using the same machinery, but 
unfortunately we have not succeeded in establishing the solution. This may be 
because the symmetry broken by distinct interactions means that there are 
insufficient kernel equations to find a full solution. This provides us with an 
opportunity to explore the limits of application of the obstinate kernel method.
\ack
Financial support from the Australian Research Council via its Discovery 
Projects scheme and the Centre of Excellence for Mathematics and Statistics of 
Complex Systems is gratefully acknowledged by RT and AO. 
Financial support from the Natural Sciences and Engineering Research Council of 
Canada via its Discovery Grant is gratefully acknowledged by AR. 
RT, acknowledges financial support from the University of 
Melbourne via its Melbourne Research Scholarships scheme. Additionally, RT and 
AO  thank the Department of Mathematics, University of British Columbia, for 
hospitality. 

\appendix
\section{$J(c,z)$: Leading coefficient of the differential equation}
\label{appendix:3fw_g1c_DE_leading_coeff}
The following is the leading polynomial coefficient of the linear homogeneous 
differential equation \eref{3fw_final_short_DE_J} satisfied by the generating 
function $J(c;z)$.

{\tiny
\begin{equation}
\fl
\eqalign {
&-2 (-1+c)^{15} \left(2-10 c+5 c^2\right) z^3-(-1+c)^{13} \left(-39+161 c+5 c^2-222 c^3+100 c^4+10 c^5\right) z^4 \\
&+(-1+c)^{12} \left(37+868 c-4988 c^2+6268 c^3-2741 c^4+1048 c^5-894 c^6+276 c^7\right) z^5 \\
&-(-1+c)^{11} \left(144-2972 c+5580 c^2+25430 c^3-54470 c^4+30904 c^5-6709 c^6+5072 c^7-2974 c^8+340 c^9\right) z^6 \\
&+(-1+c)^{10} \left(64-4392 c+50474 c^2-199461 c^3+206342 c^4-40697 c^5+80412 c^6-165265 c^7+79458 c^8-4196 c^9-2640 c^{10}+144 c^{11}\right) z^7 \\
&+(-1+c)^9 c \left(32+1124 c+196538 c^2-1302168 c^3+2311239 c^4-1877980 c^5+1680410 c^6-1880689 c^7+1132844 c^8 \right . \\
&- \left . 267138 c^9+11452 c^{10}+1008 c^{11}\right) z^8 \\
&-(-1+c)^8 c \left(-768+35664 c-760080 c^2+2816159 c^3-2310559 c^4-765330 c^5-299959 c^6+349759 c^7+4145937 c^8 \right . \\
& \left . -4922524 c^9+1924892 c^{10}-255032 c^{11}+5616 c^{12}\right) z^9 \\
&-2 (-1+c)^7 c^2 \left(2272+323820 c-4500222 c^2+19063995 c^3-37596741 c^4+48558131 c^5-56624432 c^6+52032149 c^7-25304076 c^8 \right.\\
&\left . +2280008 c^9+2360242 c^{10}-676758 c^{11}+49032 c^{12}\right) z^{10}\\
&-(-1+c)^6 c^2 \left(44544-230784 c+3551112 c^2-38087632 c^3+180802288 c^4-453709471 c^5+757037039 c^6 \right. \\
& \left . -984837233 c^7+964461909 c^8 -610442720 c^9+210975064 c^{10}-28939008 c^{11}-1107832 c^{12}+435024 c^{13}\right) z^{11}\\
&+(-1+c)^5 c^2 \left(70656-986624 c+3821136 c^2-835488 c^3+18490772 c^4-384953050 c^5+1637076179 c^6 \right. \\
& \left.-3696376911 c^7+5534602531 c^8 -5744764453 c^9+3949902310 c^{10}-1648705682 c^{11}+364273136 c^{12}-32000516 c^{13}+400464 c^{14}\right) z^{12}\\
&+2 (-1+c)^4 c^2 \left(-12288+486912 c-7890144 c^2+47462076 c^3-120173060 c^4+146708495 c^5-604330390 c^6+2911050330 c^7-7397407941 c^8 \right. \\
& \left. +11555022726 c^9-12066613597 c^{10}+8462237673 c^{11}-3794267461 c^{12}+989457534 c^{13}-128435640 c^{14}+6705720 c^{15}\right) z^{13}\\
&+(-1+c)^4 c^3 \left(270336-17143296 c+205053760 c^2-1034863488 c^3+2602602184 c^4-3573368080 c^5+5348902942 c^6-13198755668 c^7 \right. \\
&\left. +25758949628 c^8-33181320397 c^9+28810239411 c^{10}-16430320530 c^{11}+5648981962 c^{12}-1011435820 c^{13}+72248976 c^{14}\right) z^{14}\\
&+(-1+c)^3 c^4 \left(8306688-167047680 c+1173463616 c^2-4823571904 c^3+11089729840 c^4-12279891800 c^5+3293103356 c^6+4133511414 c^7 \right. \\
&\left. +2455498351 c^8-18471969408 c^9+28896185625 c^{10}-24138273334 c^{11}+11229185308 c^{12}-2621954160 c^{13}+223736688 c^{14}\right) z^{15}\\
&+(-1+c)^3 c^4 \left(1572864-51683328 c+155830784 c^2+141084224 c^3-528092320 c^4+5567010008 c^5-32548560748 c^6+83947849496 c^7 \right.\\
&\left.-126563190548 c^8+123834766557 c^9-73795306179 c^{10}+18217543880 c^{11}+4671069798 c^{12}-3521414796 c^{13}+470571408 c^{14}\right) z^{16}\\
&+(-1+c)^2 c^5 \left(4227072+295956480 c-4920437248 c^2+25943305984 c^3-72027569912 c^4+141762670800 c^5-245329809174 c^6 \right. \\
&\left. +384940046996 c^7-502611131497 c^8+500225586868 c^9-345499267041 c^{10}+144810235308 c^{11}-27350454036 c^{12}-978635976 c^{13}+735194736 c^{14}\right) z^{17}\\
&+2 (-1+c)^2 c^6 \left(59228160-1778718720 c+14159250944 c^2-51407058304 c^3+108600982392 c^4-162102121468 c^5+207049334756 c^6 \right. \\
& \left. -255318789184 c^7+290088693149 c^8-257066824433 c^9+148883529230 c^{10}-46264242442 c^{11}+4624958074 c^{12}+472617288 c^{13}\right) z^{18}\\
&+8 (-1+c)^2 c^7 \left(-101904384+1601235456 c-8594714752 c^2+23062614784 c^3-37097520808 c^4+41425091734 c^5 \right. \\
&\left. -40121948820 c^6+45641602984 c^7-53849236926 c^8+43926256959 c^9-19045148173 c^{10}+3012130047 c^{11}+140431050 c^{12}\right) z^{19} \\
&+4 (-1+c) c^8 \left(-554409984+6245716992 c-27676186624 c^2+66087476672 c^3-95061235200 c^4+84487121352 c^5-54752814068 c^6 \right. \\
& \left. +65552043352 c^7-109024512354 c^8+105526411092 c^9-49565408243 c^{10}+8416540621 c^{11}+319339692 c^{12}\right) z^{20} \\
&+16 (-1+c) c^9 \left(189984768-1521400320 c+5054821760 c^2-8837982112 c^3+7063787880 c^4+1606610060 c^5-5197447550 c^6 \right. \\
&\left. -4952167254 c^7+14081942129 c^8-9650484773 c^9+2082577819 c^{10}+79382673 c^{11}\right) z^{21}\\
&+16 c^{10} \left(121503744-774825984 c+2029717248 c^2-1682295616 c^3-4520050192 c^4+14347256304 c^5-14597362124 c^6+1236304748 c^7 \right. \\
&\left. +9045315224 c^8-6669607380 c^9+1402412613 c^{10}+61625151 c^{11}\right) z^{22}\\
&+128 c^{11} \left(12288-15366144 c+119389120 c^2-484213744 c^3+1107002264 c^4-1349066260 c^5+672939978 c^6+180870952 c^7 \right. \\
&\left. -322549025 c^8+86688397 c^9+4300398 c^{10}\right) z^{23}\\
&+128 c^{12} \left(-5640192+39028224 c-152515968 c^2+372571392 c^3-525698384 c^4+368110728 c^5-48271780 c^6 \right. \\
&\left. -78212050 c^7+29016170 c^8+1603611 c^9\right) z^{24}\\
&+1024 c^{13} \left(285696-2133888 c+7700544 c^2-13825904 c^3+12138088 c^4-3739140 c^5-1217146 c^6+740346 c^7+44559 c^8\right) z^{25}\\
&+16384 c^{15} \left(23808-73632 c+82960 c^2-35756 c^3-2310 c^4+4358 c^5+279 c^6\right) z^{26}
}
\end{equation}
}

\section*{References}


\end{document}